\documentclass[twocolumn,prb,amsmath,amssymb]{revtex4}

\usepackage{graphicx}
\usepackage{dcolumn}
\usepackage{bm}
\usepackage{color}
\usepackage{natbib}
\usepackage[titletoc,toc,title]{appendix}
\usepackage{wrapfig}

\begin{document}



\title{\large \bf Feynman Path Integral Approach on Superconducting  Qubits and Readout Process}

\author{Ali Izadi Rad, Hesam Zandi, Mehdi Fardmanesh }
\affiliation{School of Electrical Engineering, Sharif University of Technology, Tehran, Iran\\
Superconductor Electronics Research Laboratory (SERL)}
\email{izadirad@ee.sharif.edu}

\date{March 21, 2014}

\thispagestyle{empty}

\begin{abstract}

In this paper we introduce a new procedure on precise analysis of various physical manifestations in superconducting Qubits using the concept of Feynman path integral in qunatum mechanics and quantum field theory. Three specific problem are discussed, we devote the main  efforts to studying the wave-function and  imaginary part of the energy of the pseudo-ground state of the Hamiltonian in Phase Qubits and we estimate  decay rate, and thus the life time of meta-stable states using the approach of 't Hooft's Instantons model.  Correction to the Tilted-Washboard potential and Current of Phase Qubits by precise analysis of Ginzburg-Landau's free energy equation has been considered. Also we evaluate the most accurate value of  energy levels and wave-functions in Charge and Flux Qubits by semi-classical approximation in path integral formalism by considering  limits of experimental errors, comparing them with WKB results and  finally, we try to study more specific the evolution of spectrum of Hamiltonian in time after addition of  interaction Hamiltonian,  in order to obtain the  high fidelity quantum gates. 
%

\end{abstract}

\maketitle


\section{Introduction}

The field of  quantum computation and the idea of universal quantum simulator was first introduced by Richard Feynman in 1982\citep{101},\citep{102}. He showed that a classical Turing machine would presumably experience an exponential slowdown when simulating quantum phenomena, while his hypothetical universal quantum simulator would not. the algorithm for quantum computing and quantum information theory developed fast, first work done with Shor's algorithm\citep{103},\citep{104}. This algorithm can be used to quickly factories large number more fast and it also have a profound effect on cryptography, or Lov Grover's algorithm that search unsorted database faster than conventional computer\citep{105}.
 
Over the past decades, along with the theoretical development, People have tried to find the systems which suitable for using them as the component of quantum computers. The Basic buildong block of quantum computer which is the heart of quantum quantum information is a two level quantum systems known as quantum bits or Qubits. If the quantum bit locates at it's ground state, it represent the state $|0\rangle$ and if the Qubits excited,  it locates at the state $|1\rangle$.

Fortunately there are many systems that can be candidate for selecting them as two level quantum systems\citep{107,108,109,110}. Qubits which invented by using the solid-state devices satisfy what we requires from Qubits such as atoms, ions and nucleus. One of the promising candidates are devices based on superconducting Josephson junction. These devices show the quantum effects in macroscopic scale and it's the most advantage of these devices\citep{106},\citep{111}.Josephson Qubits can be categorized into three
main classes: Charge, phase and flux Qubits, and in each system one of the physical measurable quantity is our quantum parameter and suitable for manipulation  \citep{112} .

A viable quantum computer needs to has a  the stable and long-live Qubits that make their coherency for long time before the manipulation and operation acts on them.

 Thus
In order to building the quantum bits and quantum gates with high accuracy and high fidelity we need to have a deep recognition to the exact description physics of the system which in translation to quantum mechanics, it mean we should have a precise analysis on evolution of  Hamiltonian which is completely time-dependent. Finding the energy levels of qubits and their corresponding states in the time-independent regime is easiest than the time-dependent form. There are various method to solve the time-independent quantum mechanics problem such as perturbation theory or WKB theory. But in fact in fact WKB is uncontrolled approximation in general and it is hard to say that the result of this methods is accurate or not\citep{1},\citep{2}. Therefor find the methods that help us to get the more accurate and reliable result is very important and essential. In the other hand there is always thermal noises in the physical system and measurements do with certain accuracy, therefore finding the analytical parameter of system must be proportional to practical inaccuracy.in fact as we shown the preceding section, in practice we need to know the value of energy levels of order $\hbar$ .

 In this paper we claim that functional formalism of quantum mechanics and Feynman path integral give us the more accurate answer that we need, admittedly the formalism of path integral has been built completely time dependent and it’s the biggest treasure that lies down in this formalism. The basic idea of the path integral formulation and using of the Lagrangian in quantum mechanics introduced by P. A. M. Dirac and Norbert Wiener \citep{201},\citep{202} The complete method was developed in 1948 by Richard Feynman. 
Broadly speaking there are two approache  to the formulation of quantum mechanics, the operator approach based on the canonical quantization of physical observables and the associated operator algebra and Feynman path integral, in fact the path integral formulation has many advantages most of which explicitly support an intuitive understanding of quantum mechanics and this is now the standard mathematical formalism used by most physicists working in quantum field theories, because it is extremely powerful, highly visual, and will often yield answers far more quickly and simply than other formalisms. 

 Here at first we review on Path integral formalism, the we study to specific Hamiltonian and bounce motion. Then we review on Instanton model, based on path integral and apply it's result to understanding the Hamiltonian in there types of superconducting Qubits and discuss on results. Finally we discuss on time-dependency of Hamiltonian in superconducting Qubits and their measurement and it's application on building the quantum gates.
\section{Path Integral Formalism}

\subsection{Basic Idea of Path Integral}
More than any other formulation of quantum mechanics, the path integral formalism is based on connections to classical mechanics. In the classical mechanics, trajectories are allowed in configuration space which extermize action functional. A principal constraint to be imposed on any such trajectory is energy conservation. By contrast in quantum mechanics particle have a little bit more freedom than their classical counterparts. A classical particle is always reflected by a potential barrier if it’s energy is lower than the potential. In contrast, a quantum particle has a non-vanishing probability to tunnel through a barrier, a property also called barrier penetration. Also in particulare by the uncertainty principle, energy conservation can be violated by a an amount $\Delta E$ over a time $\sim \hbar/\Delta E$. The connection to action principles of classical mechanics becomes particularly apparent in problem of quantum tunneling. A particle with energy $E$ can tunnel through a potential of height $V$ which $V>E$ . However, this process is penalized by a damping factor 
$\exp{(i\int pdx/\hbar)}$ where $p=\sqrt{2m(E-V) }$ and integration is over barrier. This means that the exponent of the action associated with the classically forbidden path. These observation motivate the idea of new formulation of quantum propagation, could it be as in classical mechanics, the quantum amplitude $A$ for propagating between any two points in coordinate space is again controlled by the action functional. One might speculate that the amplitude in quantum mechanics is obtained as $A\sim \sum_{x(t)} exp(iS[x]/\hbar)$ where this sum is over all paths compatible with initial conditions of the problem and $S$ stands for the classical action. With this ansatz some feature of quantum mechanics can be observe. In the classical limit, $\hbar \rightarrow 0 $ , the amplitude would become dominated by contribution to the sum from classical path. Also quantum mechanical tunneling would be natural element of theory, non-classical path do contribute to the net amplitude but at the cost of damping factor specified by the imaginary action. By this introduction we maybe ready to constructing the formalism.

\subsection{Construction of Path Integral}
In this section we discuss on formalism of Feynman path integral  in general form which the Hamiltonian, $\mathbf{H}$, can be function of time. We emphasize on time dependency of functionals in theory, because this is one of the main reason that we use from Feynman path integral formalism to solving our problem.

We consider a bounded operator in Hilbert space, $U(t,t')$, $t \geq t'$ , which describes the evolution from time $t'$ to time $t$ and  satisfies a 
Markov property in time  \citep{2.0}
\begin{equation}\label{a1000}
U(t,t'')U(t'',t')=U(t,t') \ \ \ \ \ \text{for} \ \ \ \ t \geq t'' \geq t 
\end{equation}
Also we consider $U(t',t')=\mathbf{1}$.
moreover, we assume that U(t,t') is differentiable with a continues derivative. We set 
\begin{equation}
\label{a1}
\frac{\partial U(t,t')}{\partial t}|_{t=t'}=-\frac{H(t)}{\hbar}
\end{equation}
here $\hbar$ is real parameter and as we know later it becomes Planck's constant. With this two fundamental properties we can obtain interesting result. By differentiating the Eq.\ref{a1000} with respect to t and take the $t''=t$ we find 
\begin{equation}\label{a2}
\hbar \frac{\partial U}{\partial t}(t,t')=-H(t)U(t,t')
\end{equation}

In the simple case, if $\mathbf{H}$ be time-independent the propagator has the simple form $U(t,t')=e^{-\frac{i(t-t')H}{\hbar}}$ and $U$ become additive, $U(\Delta t_1)U(\Delta t_2)=U(\Delta t_1+\Delta t_2)$. Markov property allow us to write 

\begin{equation}
U(t'',t')= \prod_{m=1}^{n} U[t'+m\epsilon,t'+(m-1)\epsilon] , \ \ \ \ \ \ n\epsilon =t''-t'
\end{equation}
Which it means propagating from certain time to the other time is equals to product of subpropagtion in tiny times. From property of Hilbert space we know 
$\mathbf{q}|q\rangle =q |q \rangle , \langle q'|q\rangle =\delta(q-q') ,  \int dq |q\rangle \langle q| =\mathbf{I}$. Then 
by defining $ t_k=t'+k\epsilon, q_0=q' , q_n=q''$ we can represent the propagator in the basis of space 

\begin{equation}\label{a7}
\langle q''|U(t'',t')|q'\rangle = \int \prod_{k=1}^{n-1} dq_{k} \prod_{k=1}^{n} \langle q_k |U(t_k,t_{k-1})|q_{k-1} \rangle 
\end{equation}

The most general Hamiltonian with keeping the time-dependency can write is 
\begin{equation}\label{a10}
\mathbf{H(t)}=\frac{{\mathbf{P}}^2}{2m}+V(\mathbf{q},t)
\end{equation}
By using the Eq. \ref{a2}, we have 
\begin{equation}\label{a11}
-\hbar \frac{\partial }{\partial t}\langle q|U(t,t')|q'\rangle =[-\frac{{\hbar}^2}{2m}{\nabla}_q^2+V(q,t)]\langle q|U(t,t')|q'\rangle 
\end{equation}

%
to solve Eq. \ref{a11} in the limit $t-t' \rightarrow 0$ it is convenient to set 

\begin{equation}
\langle q|U(t,t')|q'\rangle =e^{-\frac{\sigma(q,q';t,t')}{\hbar}}
\end{equation}

Eq. \ref{a11} then becomes 
\begin{equation}\label{a111}
\partial_{t} \sigma =-\frac{1}{2m}({\nabla}_q \sigma)^2+\frac{\hbar}{2m}{\nabla}_{q}^2 \sigma +V(q,t)
\end{equation}

In the case $V=0$ it is easy from direct solution of schrodinger it is easy to observe that the solution is singular for $|t-t'|\rightarrow 0$, by using this hint, therefore it is convenient to separate terms which produce singularity and redefine $\sigma$ as 
\begin{equation}\label{a12}
\sigma(q,q';t,t')=m\frac{(q-q')^2}{2(t-t')}+\frac{d}{2}\hbar \ln [\frac{2\pi \hbar(t-t')}{m}]+\sigma_1(q,q';t,t')
\end{equation} 
Therefore 
\begin{eqnarray}
\partial_{t}\sigma &&=-\frac{m}{2(t-t')^2}(q-q')^2+\frac{\hbar d}{2(t-t')}+\partial_{t}\sigma_1 \nonumber\\
{\nabla}_q{\sigma} &&=\frac{m}{t-t'}(q-q')+{\nabla}_{q} {\sigma}_1  \nonumber\\
{\nabla}_{q}^2 \sigma &&=\frac{dm}{t-t'}+{\nabla}_{q}^2 \sigma_1
\end{eqnarray}

By putting them in equation \ref{a111} and keeping the leading term, which are order $|t-t'|$ , reduce to 
\begin{equation}
\partial_{t} {\sigma}_1=-\frac{1}{t-t'}(q-q')\cdot {\nabla}_{q} {\sigma}_1+V(q,t)
\end{equation}
It is more suggestive to write the equation as 
\begin{equation}
[(t-t'){\partial}_t+(q-q')\cdot {\nabla}_{q}]{\sigma}_1=(t-t')V(q,t)
\end{equation}

Under this form, one realize that the equation can be solves by introducing a dilation parameter $\lambda$ and function $\phi(\lambda)$ obtained by substituting $ t \mapsto t'+\lambda(t-t'), \mathbf{q \mapsto q'+ \lambda(q-q')}$ in $\sigma$ . The function satisfies the equation 
\begin{equation}
\lambda \frac{d\phi}{d\lambda}=\lambda(t-t')V(q'+\lambda(q-q'), t'+\lambda(t-t'))
\end{equation}

Therefore 
\begin{equation}
{\sigma}_1=\phi (1)=(t-t') \int_{0}^{1} d\lambda v(q'+\lambda (q-q'), t'+\lambda(t-t'))
\end{equation}

By substituing the whole terms we have 

\begin{equation}
\langle q| U(t,t')|q' \rangle =\bigg(\frac{m}{2\pi \hbar (t-t')}\bigg)^{\frac{d}{2}}e^{-\frac{S(q)}{\hbar}}
\end{equation}
Which $S(q)$ is our action and define as 
\begin{equation}
S(q)=\int_{t'}^{t} d\tau [\frac{1}{2}m{\dot{q}}^2(\tau)+ V(q(\tau),\tau)] \ \ \ +O((t-t')^2)
\end{equation}

{\bf{Remarks}} \citep{2.0}. In the expansion of $\sigma $, the most singular term for $\epsilon=t-t' \rightarrow 0 $ is $\frac{(q-q')^2}{\epsilon}$
 , which is independent of the potential. This implies that the support of the matrix element $\langle q |U(t'+\epsilon,t')|q' \rangle$ corresponds to values $|q'-q|=O(\sqrt{\epsilon})$, moreover, for $|q'-q|=O(\sqrt{\epsilon})$, one finds
\begin{eqnarray}
&&\sigma_1(q,q';t'+\epsilon, t' )=\epsilon(\frac{q+q'}{2},t')+O({\epsilon}^2) \nonumber\\
&&=\frac{\epsilon (V(q,t)+V(q',t'))}{2}+O({\epsilon}^2)\nonumber \\
&&=\epsilon V(q,t)+O({\epsilon}^{\frac{3}{2}})
\end{eqnarray} 

Now combining the short time evaluation with Eq. \ref{a7}, by introducing $\epsilon=\frac{t''-t'}{n}$ we can find 
\begin{equation}
\langle q''|U(t'',t')|q' \rangle =\lim_{n\rightarrow \infty} (\frac{m}{2\pi \hbar \epsilon})^{\frac{nd}{2}} \int \prod_{k=1}^{n-1} d^d q_k 
e^{-\frac{S(q,\epsilon)}{\hbar}}
\end{equation}

With 
\begin{equation}
S(q,\epsilon)=\sum_{k=0}^{n-1} \int_{t_k}^{t_{k+1}} dt [\frac{1}{2}m{\dot{q}}^2(t)+V(q(t),t)]+O({\epsilon}^2)
\end{equation}
Summation of whole integral yields 
\begin{equation}
S(q,\epsilon)=\int_{t'}^{t''} dt [\frac{1}{2}m {\dot{q}}^2(t)+ V(q(t),t)]+O(n {\epsilon}^2)
\end{equation}

The terms neglected in expression are of order $n{\epsilon}^2$ and when set condition $\epsilon \rightarrow 0 , n \rightarrow \infty , n\epsilon =t''-t'$ then we find that $n{\epsilon}^2=\epsilon \times (t''-t') \rightarrow 0$ and therefor this term vanish certainly.
When $\epsilon \rightarrow 0$ , the limit of $S(q,\epsilon)$ is the {\it{\bf euclidean}} action 
\begin{equation}
S(q):=\int_{t'}^{t''} dt [\frac{1}{2} m {\dot{q}}^2(t)+V(q(t),t)]
\end{equation}
{ in concolution } the matrix element can be written as 
\begin{equation}
\langle q''|U(t'',t')|q' \rangle = \int_{q(t')=q'}^{q(t'')=q''} [Dq(t)] e^{-\frac{S(q)}{\hbar}}
\end{equation}
which $Dq=\chi\lim_{N\rightarrow \infty }\prod dq_n$ reperesent a functional result resulting from taking the continuum limit of some 
finite dimensional space and th action convert to the functional , S[x].the functional resulting from taking the limit of infinitely many discretization points , $\{x_n\} $ is denoted by $x:x\mapsto x(t)$ whrer $t$ plays the role of the formerly discrete index $n$ and $\chi=( \frac{m}{2\pi \hbar \epsilon } )^{\frac{dn}{2}}$ which diverge in the limit of $n$ to infinity \citep{2}.


 \section{Semiclassical Approximation}\label{sec3}
\subsection{Expansion Around Classical Path}
The points of stationary phase is a  configuration $\bar{x}$, which  qualified by condition of vanishing functional derivative 
\begin{equation}
D F_x=0 \Longleftrightarrow \forall t:\frac{F[x]}{\delta x(t)}|_{x=\bar{x}}=0
\end{equation}
In fact they are the classical answer which can be obtain in classical mechanic and as we now from principle of least action the classical answer are saddle point of action . In addition, the Taylor expantion ot the the functional yields 
\begin{equation}
F[x]=F[\bar{x}+y]=F[\bar{x}]+\frac{1}{2} \int dt \int dt' y(t')A(t,t')y(t) + \cdots 
\end{equation}
Where 
\begin{equation}
A(t,t')=\frac{{\delta}^2 F[x]}{\delta x(t) \delta x(t')}|_{x=\bar{x}}
\end{equation}

This kind of representation of functional and using the approximation in order of two, by using the Gaussian integral we can calculate the matrix element with approximation. This approximation is called semi-classical approximation and it means we evaluate the path which is near to classical path. With method which has been described in Appendix A we can find that 
$\int Dx \exp(-F[x]) \simeq \exp(-F[\bar{x}]) [det(A/2\pi)]^{-\frac{1}{2}}$.we expand this notion more here .

If we expand the action around the saddle points by defining $x_c$ as answer of classical path. We introduce $x(t)=x_c(t)+r(t)$ then 
\begin{eqnarray}
 S(x)&&=\int_{-\tau/ 2}^{\tau/2} [\frac{1}{2}{\dot{x_c}}^2+\frac{1}{2}{\dot{r}}^2+ V(x_c+r)]dt \nonumber \\
&& =\int_{-\tau/2}^{+\tau/2} [\frac{1}{2} {\dot{x_c}}^2(t)+V(x_c(t))]dt  \\
&&+\int_{-\tau/2}^{\tau/2} [\frac{1}{2}{\dot{r}}^2(t)+\frac{1}{2}V''(x_c(t))r(t)^2]dt \nonumber 
+O(r^3)
\end{eqnarray}

The quadratic from in $r$ can be wrriten as 
\begin{eqnarray}
\Sigma(r)&&= \int_{-\frac{\tau}{2}}^{\frac{\tau}{2}} dt [\frac{1}{2} {\dot{r}}^2(r)+\frac{1}{2}V''(x_c(t))r^2(t)]\\
&&=\frac{1}{2} \int dt_1dt_2 r(t_1) M(t_1,t_2)r(t_2)
\end{eqnarray}

Where 
\begin{equation}
M(t_1,t_2):=\frac{{\delta}^2 S}{\delta x_c(t_1) \delta x_c(t_2)}=[-d_{t_1}^2+V''(x_c(t_1),t)]\delta(t_1-t_2)
\end{equation}
the differential operator $M$ act on a function $r(t)$ as the form of a Hermitian quantum Hamiltonian, t playing the role of a position variable and $V''(x_c(t))$ being the potential . All its eigenvalue are real as well as its determinant.

\subsection{Zero mode}

The classical path, $x_c$ can be estimate by Euler-Lagrange equation for Euclidean Lagrange which yields $-{\ddot{x}}_c(t)+V'(x_c,t)=0$ therefore $\dot{x}_c$ is an Eigen function of operator $M$ whit eigenvalue of zero which called zero mode 
\begin{equation}
[-d_{t}^2 +V''(x_c)]\dot{x_c}
\end{equation}
Therefor $[\text{det M}]^{-\frac{1}{2}} $ becomes infinite .in the next part we discusse on this problem\citep{2.0}.

\subsection{Collective Coordinates}

In the case of a path integral, it is necessary to integrate exactly over the variables that parameterize the saddle points, the so-called collective coordinate \citep{22}-\citep{23}.
We start from the identity 
\begin{equation}
1=\frac{1}{\sqrt{2\pi \xi}}\int_{-\infty}^{\infty} d\lambda e^{-\frac{{\lambda}^2}{2\xi}}
\end{equation}
Where $\xi$ is an arbitrary constant. we then change variables, $\lambda \mapsto t_0$ with 

\begin{equation}
\lambda :=\int dt \dot{x}_c(t)(x(t+t_0)-x_c(t))
\end{equation}

We obtain a new identity
\begin{equation}
1=\frac{1}{\sqrt{2\pi \xi}}\int dt_0 [\int dt \dot{x}_c(t) \dot{x}(t+t_0)]e^{-\frac{1}{2\xi}[\int dt \dot{x_c}(t)(x(t+t_0)-x_c(t))]^2}
\end{equation}

Now we put this integral in the path integral equation and it yields the modified action 
\begin{equation}
S_{\xi}(x)=S(x)+\frac{\hbar}{2\xi }[\int dt \dot{x_c}(t) (x(t+t_0)-x_c(t))]^2
\end{equation}
if we estimate the new saddle point of the action 
\begin{equation}
\frac{\delta S}{\delta x(t)}+\frac{\hbar }{\xi }\dot{x_c}(t)\int dt' [\dot{x}_c(t')-x_c(t')]=0
\end{equation}
We can see the answer of this equation is again is $x=x_c(t)$. now by estimating the second derivation of $S$ we have 

\begin{equation}
\frac{{\delta}^2 S}{\delta x_c(t_1) \delta x_c(t_2)} \rightarrow M_{\xi}(t_1,t_2) := \frac{{\delta}^2 S}{\delta x_c(t_1) \delta x_c(t_2)}+\frac{\hbar}{\xi}\dot{x_c}(t_1)\dot{x_c}(t_2)
\end{equation}

By this new matrix $M$ we don't get zero eigenvalue for $\dot{x}_c(t)$ and instead we have eigenvalue $\mu=\frac{\hbar ||\dot{x_c}||^2}{\xi}$

if we define 
\begin{equation}
M_0(t_1,t_2)=[-d_{t_1}^2+1]\delta(t_1-t_2)
\end{equation}
then 
\begin{eqnarray}
\text{det} M_{\xi}&&=\text{det}(M+\mu |0\rangle \langle 0|)M_{0}^{-1}\\
&&=\lim_{\epsilon\rightarrow 0}\text{det} (M+\epsilon +\mu |0\rangle \langle 0|)(M_0+\epsilon)^{-1}
\end{eqnarray}

we define 
\begin{equation}
\lim_{\epsilon \rightarrow 0}\frac{1}{\epsilon}\text{det}(M+\epsilon)(M_0+\epsilon)^{-1}:=\text{det}' MM_0^{-1}
\end{equation}
it is easy to show that 
\begin{equation}
\text{det}(M+\epsilon +\mu |0\rangle \langle 0|)(M_0+\epsilon)^{-1}=(1+\frac{\mu}{\epsilon})\text{det}(M+\epsilon)(M_0+\epsilon)^{-1}
\end{equation}
and in the limite  $\epsilon \rightarrow 0$ it becomes 

\begin{equation}
\text{det}'MM_{0}^{-1}||\dot{x_c}||^2\frac{\hbar}{\xi}
\end{equation}

Collecting all factors, one concludes that the Gaussian integration over configuration in the neighboring of saddle point yields a factor \citep{2.0}

\begin{equation}
\frac{\tau}{\sqrt{2\pi \hbar}}||\dot{x_c}||(det' MM_0^{-1})^{-\frac{1}{2}}e^{-\frac{\tau}{2}}
\end{equation}

\subsection{Collective Coordinate: Alternative Method}

If $q_c(t)$ be the particular solution of the saddle point equation of action that corresponding to $t_0=0$ , thus the general solution is $q_c(t-t_0)$ . To introduce an integration variable associated with time translation we set 
\begin{equation}
q(t)=q_c(t-t_0)+r(t-t_0) \sqrt{\hbar}
\end{equation}

Here we have to degree of freedom and we should reduce it to on freedom, thus we add on restriction such as 
\begin{equation}
\int \dot{q}_c(t-t_0)r(t-t_0) dt =0 
\end{equation}

We then expand $r(t)$ on the orthonormal basis formed by the eigenvectors $f_n(t)$ of the Hermitian operator
\begin{equation}
r(t)=\sum_{n=0}^{\infty} r_n f_n(t)
\end{equation}

The set $\{ f_n \}$ includes the normalized eigenvector $ \frac{\dot{q}_c}{||\dot{q}_c||}$ , which we can identify with $f_0$ . We see that $r_0$ of $f_0$ vanish .therefor
\begin{equation}
\frac{1}{2} \int dt_1 dt_2 r(t_1) M(t_1,t_2) r(t_2) =\frac{1}{2}\sum_{n>0} m_n {r_n}^2
\end{equation}
Where $\{ m_n \}$ is the set of all non-vanishing eigenvalues of $M$ . If we extend the the $q(t)$ on an orthonormal basis of square-integrable functions 
\begin{equation}
q(t)=\sum_{m=0}^{\infty} c_m g_m(t) , \ \ \ \ \ \ g_m(t) \in \mathbf{L^2}
\end{equation}

Then 
\begin{equation}
Dq=\chi \prod_{m=0}^{\infty}{dc_m}
\end{equation}

For the estimating of Jacobian because of changing the variables from $q(t)$ to set ${t_0,r_n}$ we write 

\begin{equation}
q(t)=q_c(t-t_0)+ \sum_{n=1}^{\infty} r_{n} f_{n}(t-t_0)
\end{equation}

The variable $c_m$ can be expressed in terms of the new variables

\begin{equation}
c_m=\int dt g_m(t) q_c(t-t_0) + \sum_{n=1}^{\infty} r_n \int dt g_m(t) f_n(t-t_0) 
\end{equation}
And clearly the Jacobian of the transformation is the determinant of the matrix 

\begin{equation}
[\frac{\partial c_m}{\partial t_0}, \frac{\partial c_m}{\partial r_n }]
\end{equation}

With 
\begin{equation}
\frac{\partial c_m }{\partial t_0}=-\int dt g_m(t) {\dot{ q}}_c(t-t_0)-\sum_{n=1}^{\infty} r_n \int dt g_m(t) {\dot{f}}_n(t-t_0) 
\end{equation}
And 
\begin{equation}
\frac{\partial c_m}{\partial r_n}=\int dt g_m(t) f_n(t-t_0)
\end{equation}

In a leading order calculation, one can neglect the dependence of the Jacobin on the ${r_n}$. Since the set $ \{\frac{{\dot{q}}_c}{||\dot{q}_c|},f_n \}$ forms an orthonormal basis, the matrix 
\begin{equation}
[\int dt g_m(t) \frac{{\dot{q}}_c(t-t_0)}{||{\dot{q}}_c||} , \int dt g_m(t) f_n(t-t_0)]
\end{equation}
Is orthogonal and it's determinant is 1. Thus the Jacobian of transformation on leading order is 
\begin{equation}
||{\dot{q}}_c||=[\int dt {\dot{q}}_c^2(t)]^{\frac{1}{2}}
\end{equation}
\subsection{Application in Superconducting Qubits}
Now we are ready to take a look on superconducting Qubits and try to analyse the spectrum of their Hamiltonian. Here we try to find the ground state of flux Qubit which it has been introduce in the next sections.

The exact form of Hamiltonian for flux Qubit given by 
\begin{equation}
V(\phi)=E_J(1-\cos{\phi})+E_L \frac{(\phi-{\phi}_e)^2}{2} \nonumber \\
\end{equation}
if we expand the cosine term by approximation the potential is near to quartic potential. For simplicity in calculation lets to analyze  the potential that has the form
\begin{equation}
H=\frac{1}{2}p^2+\frac{1}{2}(x^2-\frac{1}{4})^2
\end{equation} 
{\bf Remarks}. Our potential can computed by path integral exactly but for simplicity we neglect from higher order of expansion, also we try to solve this problem exactly using the Instanton model as alternative way in preceding sections. In the appendix C we solve this problem with WKB method too.

Now, at first we try to find the classical path, by extermizing the action we have 
\begin{equation}
-\ddot{x}_c(t)+2x_c(t)[x_c^2(t)-\frac{1}{4}]=0
\end{equation}
Therefore the classical path given by 
\begin{equation}
x_c(t)=\pm \frac{1}{2} \tanh(\frac{t-t_0}{2})
\end{equation} 
Where $t_0$ in constant of integration.by calculating the action for classical path we have 
\begin{eqnarray}
S(x_c) &&=\lim_{\tau \rightarrow \infty} \int_{-\tau/2}^{\tau/2}[\frac{1}{2}{\dot{x}(t)}^2+\frac{1}{2}(x^2(t)-\frac{1}{4})^2]dt \nonumber \\
 && =\frac{1}{6}
\end{eqnarray}
In the oder hand 
\begin{eqnarray}
\text{tr} Pe^{-\frac{\tau H}{\hbar}} &&= \sum_{\pm, n} \pm e^{-\frac{\tau E_n,\pm}{\hbar}}  \\
&& \sim -2\sinh[\frac{\tau(E_{0,+}-E_{0,-})}{2\hbar}]e^{-\frac{\tau(E_{0,+}+E_{0,-})}{2\hbar}} \nonumber 
\end{eqnarray}
where $P$ is parity operator. Thus 
\begin{equation}
\frac{\text{tr}P e^{-\frac{\tau H}{\hbar}}}{\text{tr}e^{-\frac{\tau H}{\hbar}}}\sim -\frac{\tau}{2\hbar}(E_{0,+}-E_{0,-})
\end{equation}

Therefore in our case, it yields to \citep{2.0}
\begin{equation}
\Delta E=E_{-}-E_{+}=2\sqrt{\frac{\hbar}{\pi}}e^{-\frac{1}{6\hbar}}(1+O(\hbar))
\end{equation}

The other property of this potential has been discussed in the preceding sections. Now it is possible to study our potential by changing the variables that we have as try the same approach which we discussed here.



%
%
%


\section{Unstable States and Bounces}

We begin with situation where in the Hamiltonian 
\begin{equation}
H=\frac{1}{2m}p^2+V(q)
\end{equation}
The potential has an absolute minimum at the origin where 
\begin{equation}
V(q)=\frac{1}{2}m {\omega}^2 {q^2}+O(q^3)
\end{equation}

We provide a situation which the minimum of the potential occur at $q=0$ and there are some $q$ that $V<0$ .
Here we calculate the imaginary part of $Z(\tau/\hbar )= \text{tr} e^{-\frac{\tau H}{\hbar}}$ for $\tau\rightarrow \infty $ . The result expected to have form 
\begin{equation}
\text{Im} Z(\tau /\hbar)\sim \text{Im} e^{-\frac{\tau E_0}{\hbar}} \sim -\frac{\tau}{\hbar}\text{Im} E_0 e^{-\tau\frac{Re E_0}{\hbar}}
\end{equation}

At leading order in $\hbar$ , $Re E_0$ can be replaced by the value it's assume in the harmonic approximation and thus 
\begin{equation}
\text{Im} Z(\tau/\hbar)\sim -\frac{\omega \tau}{\hbar} e^{-\frac{\omega \tau}{2}}\text{Im} E_0
\end{equation}

Now, we look for non trivial saddle points of the path integral. The saddle point equation obtained by varying the Euclidean action with boundary equation
\begin{equation}
-m{\ddot{q}}+V'(q)=0 , \ \ \ \ \ \\ q(-\frac{\tau}{2})=q(\frac{\tau}{2})
\end{equation}

This equation lead to constant of motion or energy by integrating from equation 

\begin{equation}
\frac{1}{2}m {\dot{q}}^2-V(q)=\epsilon , \epsilon <0 
\end{equation}
If we name the two point that the velocity of classical particle go to zero by $ q_{+} $ and $ q_{-}$ then for one period 
\begin{equation}
\tau= 2 \sqrt{m} \int_{q_{-}}^{q_{+}} \frac{dq }{\sqrt{2V(q)+2\epsilon}}
\end{equation}

The period $\tau$ diverge only for constants $\epsilon$ such that $V(q)+\epsilon$ has double zero at $q_{-}$ or $q_{+} $ and it implies that $V'(q)$ vanish. Moreover the action remains finite in this limit only if $V(q(t))$ and $ \dot{q}$ vanish for $|t| \rightarrow \infty $ these conditions are compatible only if $\epsilon$ and thus $q_{-}$ vanish


\subsection{Simple Example }

Here we consider the simple case to describe the detail of calculation and in the next sections we develop this idea to the general from and our main problem. 
we consider the quartic anharmonic potential in which the sign of the quartic term is changed from positive to negative value. The corresponding Hamiltonian is 
\begin{equation}
H=\frac{1}{2}{\dot{q}}^2(t)+ \frac{1}{2}q^2(t)+\frac{1}{4}g q^4(t) 
\end{equation}
the factor $g$ is coupling factor and play the main rule in our calculation.in order to analyze the transition rate and states of energy we need to estimate the partition function. We consider $\hbar=1$ 
\begin{equation}
Z(\beta)=tr e^{-\beta H}=\int_{q(-\beta/2)=q(\beta/2)} Dq e^{-S(q)}
\end{equation}
And $S(q)$ is an Euclidean action 
\begin{equation}
S(q)=\int_{-\beta/2}^{\beta/2}dt [\frac{1}{2}{\dot{q}}^2(t)+\frac{1}{2}q^2(t)+\frac{1}{4}g q^4(t)]
\end{equation}

for the domain which $Re(g)>0$ , the integral is analytical and has a finite answer.following the steepest descent method, the integral is dominated for $g\rightarrow 0$ by saddle point $q(t)=0$ .thus it can be calculated by expanding the integrand in power of $g$ and integrating the successive terms. This leads to perturbing the partition function by using Appendix B.

but the more interesting case where have more effect on the final result are the the cases which the sign of $g$ is negative .for all $g<0$ the Hamiltonian is no longer bonded from below therefore the energy eigenvalues, considered as analytics function of $g$ have a singularity at $g=0$ and the perturbation diverge. Following example describe the challenge .

If we consider this integral 
\begin{equation}
\Lambda(g):=\frac{1}{\sqrt{2\pi}}\int_{-\infty}^{\infty} e^{-[\frac{x^2}{2}+g\frac{x^4}{4}]}dx
\end{equation}

For positive $g$ we can estimate it by steepest descent method and write it as 
\begin{equation}
\Lambda(g)=1+O(g)
\end{equation}
and the function $\Lambda$ is analytical in a cut plane.to continue the integral analytically to $g<0$ it is necessary to rotate the integration contour $C$ when one changes the phase of $g$ , for example like $C: \text{Arg} \ \ x=-\frac{1}{4}\text{Arg} \ \ g $ then always $\text{Re}(gx^4) $ remains positive. We set two contours 
\begin{eqnarray*}
\text{for} \,g=-|g|+i0 &&: {\Lambda}_{+}(g)=\int_{C_{+}} \frac{e^{-(\frac{x^2}{2}+g \frac{x^4}{4})}}{\sqrt{2\pi}}dx\\
\text{with } C_{+} &&: \text{Arg}\, x=-\frac{\pi}{4}\\
\text{for} \,g=-|g|-i0 &&: {\Lambda}_{-}(g)=\int_{C_{-}} \frac{e^{-(\frac{x^2}{2}+g \frac{x^4}{4})}}{\sqrt{2\pi}}dx\\
\text{with } C_{-} &&: \text{Arg}\, x=\frac{\pi}{4}
\end{eqnarray*}

For $g \rightarrow 0_{-}$ the two integrals are still dominated by the saddle point at the origin since the contributions of the other saddle points
\begin{equation}
x+gx^3=0 \rightarrow x^2=-\frac{1}{g}
\end{equation}
Therefor it is of order 
\begin{equation}
e^{-(\frac{x^2}{2}+\frac{gx^4}{4})}\sim e^{\frac{1}{g}} \ll 1
\end{equation}

However the discontinuity of $\Lambda(g)$ across the cut is given by the difference between the two integrals
\begin{equation}
2i \text{Im}\, \Lambda(g)=\Lambda_{+}(g)-\Lambda_{-}(g)=\frac{1}{\sqrt{2\pi}}\int_{C_{+}-C_{-}}e^{-(\frac{x^2}{2}+g\frac{x^4}{4})}dx
\end{equation}

The contribution of the saddle points at the origin cancel and the non-trivial saddle point at $x=\pm 1/\sqrt{-g}$ remain which by integrating on the contour $C_{+}-C_{-}$ we have 
\begin{equation}
\text{Im} \, \Lambda(g) \sim 2^{-\frac{1}{2} }e^{\frac{1}{4g}}
\end{equation}

as a consequence, for $ g$ negative and small while the real part of the integral is given by the perturbation expansion the leading contributions to the imaginary part, which is exponentially small,come from the non-trivial saddle points. Now if we apply this calculation to path integral we should be carefull.

The domain must satisfy $\text{Re}({\dot{q}}^2(t))>0$. therefore for negative $g$ there are two condition
\begin{equation}
\text{Re}[gq^4(t)]>0 , \ \ \text{Re}[{\dot{q}}^2(t)]>0
\end{equation} 
And they satisfy if the domain be $ \text{Arg}\, q(t)=-\theta (\text{mod} \,  \pi),\, \pi/8<\theta <\pi/4 or -\pi/4 < \theta <-\pi/8$
For $g\rightarrow 0$ the path integral corresponding to the two analytical continuations are also domain by the saddle point at the origin $q(t)=0$ . But the difference between the two integrals  contributions cancel. The other saddle point $q(t)=-1/g$ is of order $e^{\frac{\beta}{4g}}$ in it becomes negligible for $\beta \rightarrow \infty$.

Therefore we should look for non-trivial solution :
\begin{equation}
-{\ddot{q}}(t)+q(t)+gq^3(t)=0
\end{equation}
With boundary condition
\begin{equation}
q(-\beta/2)=q(\beta/2)
\end{equation}

From the equation of motion we have 
\begin{equation}
\frac{1}{2}{\dot{q}}^2(t)-\frac{1}{2}q^2-\frac{1}{4}g q^4=\epsilon
\end{equation}
which $\epsilon<0$ and $q_{-} ,q_{+}$ are places that velocity , $\dot{q}$ vanishes. In order to find the period we have 
\begin{equation}
\beta=2\int_{q_{-}}^{q_{+}} \frac{dq}{\sqrt{q^2+\frac{1}{2}gq^4+2\epsilon}}
\end{equation}

$\beta$ diverge only if $\epsilon$ and thus $q_{-}$ go to zero . Thus the classical solution becomes 
\begin{equation}
q_c(t)=\pm (-\frac{2}{g})^{\frac{1}{2}}\frac{1}{\cosh(t-t_0)}
\end{equation}
the classical action therefor is 
\begin{equation}
S(q_c)=-\frac{4}{3g}+O(\frac{e^{-\beta}}{g})
\end{equation}

The by using the semi-classical approximation we find that 
\begin{eqnarray}
M(t_1,t_2)&&=\frac{{\delta}^2S}{\delta q_c(t_1)\delta q_c(t_2)} \\
&&=[-(\frac{d}{dt_1})^2+1+3g{q_c}^2(t_1)]\delta(t_1-t_2) \nonumber
\end{eqnarray}
Matrix $M$ has zero mode and we should use the result of collective coordinate and we find that 
\begin{equation}
\text{Im}\, tr\, e^{-\beta H}\sim \frac{2}{2i}[\text{det}'M M_0^{-1}]^{-\frac{1}{2}} J \frac{\beta}{\sqrt{2\pi}}e^{-\frac{\beta}{2}}e^{\frac{4}{3g}}
\end{equation}
and after calculating the determinant we find that 
\begin{equation}
\text{Im} \, E_{0}=\frac{4}{\sqrt{2\pi}}\frac{e^{\frac{4}{3g}}}{\sqrt{-g}}[1+O(g)]
\end{equation}

\subsection{More Genreal Form of Hamiltonian }
In the most general form if we want to estimate the imaginary part of metastable state of Hamiltonian 
\begin{equation}
H=-\frac{1}{2}\frac{d^2}{dq^2}+\frac{1}{2}q^2+\frac{1}{2}g q^{2N} , \ \ \ \ \ g<0
\end{equation}
The solution is similar to the previous solution. The classical path can be written as \citep{2.0}

\begin{equation}
|q_c|(t)={\frac{1}{\sqrt{-q}}[\frac{1}{\cosh[(N-1)(t-t_0)]}}]^{\frac{1}{{N-1}}}
\end{equation}
Substituting the classical path in the action,yeilds 
\begin{equation}
S_c=\frac{A(N)}{(-g)^{\frac{1}{N-1}}}
\end{equation}
Which factor $A(N)$ has been defined as 
\begin{equation}
A(N)=\frac{\sqrt{\pi} \Gamma(\frac{N}{N-1})}{2\Gamma(\frac{N-1}{2N-2})}=4^{\frac{1}{N-1}}\frac{{\Gamma}^2(\frac{N}{N-1})}
{\Gamma(\frac{2N}{N-1})}
\end{equation}
Applying the semi-classical approximation as usual make an operator similar to previous calculation 
\begin{equation}
M=-{\partial}_t^2+1-\frac{N(2N-1)}{{\cosh}^2(N-1)t}
\end{equation}
Now comparing the obtained operator with Bargman-Wigner equation tell us by setting the desired coefficient $\lambda=\frac{N}{N-1}$ and $z=\frac{\sqrt{1+\epsilon}}{N-1}$ 
We will find 
\begin{equation}
\text{det}(M+\epsilon)(M_0+\epsilon)^{-1} \sim -2^{-\frac{N+1}{N-1}}A(N) \epsilon
\end{equation}
The operator $M$ has a vanishing eigenvalue and therefore we need to using the collective coordinate methods. Similar to previous calculation the Jacobian of this transformation is

\begin{equation}
J^2=\int dt {\dot{q}}^2(t)= \frac{A}{(-g)^{\frac{1}{N-1}}}
\end{equation}
Therefor the imaginary part of the energy of the metastable pseudo ground state is 
\begin{equation}
\text{Im} E(g)=C(-g)^{-\beta}e^{-\frac{A(N)}{(-g)^{\frac{1}{N-1}}}}
\end{equation}
with $ \beta =\frac{1}{2N-2}$ and $ C=\frac{2^{\frac{1}{N-1}}}{\sqrt{\pi}}$.

\subsection{Bargman-Wigner Potential and It's Eigenvalues} 

In order to estimating the eigenvalue of our problem. We can do this 
From semi-classical approximation in scattering theory we have \citep{24}
\begin{eqnarray}
\langle {\psi}_2 |S|{\psi}_1 \rangle && \sim \int dp'dp'' {\psi}_{2}^{*}(p''){\psi}_{1}(p')e^
{\frac{i}{\hbar}(\frac{t'{'p''}^2}{2m}-\frac{t'{p'}^2}{2m})}\nonumber \\
&&\times \langle t'',p''/m|U(t'',t')|t'p'/m\rangle
\end{eqnarray}
The elements of the $S$-matrix, $S_{+}$and $S_-$ that correspond to transmission and reflection, 
For the one-dimensional potential $V(q) =\frac{\lambda}{{\cosh}^2 q}$ can be obtain 
\begin{equation}
S_{+}=\frac{\Gamma(d+\alpha)\Gamma(\alpha+1-d)}{\Gamma(1+\alpha)\Gamma(\alpha)}
\end{equation}
Therefore for General form of operator 
\begin{equation}
M=-d_{t}^2+1-\frac{\lambda (\lambda +1){\omega}^2}{{\cosh{\omega t}}^2}
\end{equation}

We have 
\begin{equation}
\text{det}(M+\epsilon)(M_0+\epsilon)^{-1}=\frac{\Gamma(1+z)\Gamma(z)}{\Gamma(1+\lambda +z)\Gamma(z-\lambda)}  
\end{equation}
with $z=\frac{\sqrt{1+\epsilon}}{\omega}$
for instance, in the case $\omega=\frac{1}{2}$ and $\lambda =2$ we can say approximately $z-2 \simeq \epsilon$ and thus

\begin{equation}
\text{det}(M+\epsilon)(M_0+\epsilon)=\frac{(z-2)(z-1)}{(z+1)(z+2)}\simeq \frac{\epsilon}{12}
\end{equation}

Now we can apply this approach to the potential which we have in Qubits. For example in phase Qubits, we have 
\begin{equation}
V(q)\sim \cos{q}+\alpha q =1+\alpha q-\frac{1}{2}q^2+\frac{1}{24}q^4+\cdots
\end{equation}
therefor by reasonable approximation we are able to expand cosine term with quadratic terms and repeat the way that we described here for our problem. However we introduce more exact way in the next section by Instanton model.



\section{ 't Hooft's   Instantons, Double Potential Well and Bounces }
\subsection{Instanton Model}
In this Section we introduce the new way to study the bounce situation, also it is necessary to understanding the Instanton model for double well potential which more similar to Flux Qubits potential. If we consider the potential like Fig.1 

\begin{center}
\begin{figure}[h]\label{YaZahra03}
\includegraphics[height=35mm]{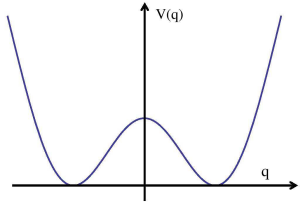}
\caption{ Double well potential which is symmetry respect to $q$, $V(q)=V(-q)$. }
\end{figure}
\end{center}



{\bf Reamrks}.the mechanism of quantum double or multiple well potential plays a role in a number of problems of condensed matter physics. For example in the Glass and amorphous solids the absence of long range force
\begin{wrapfigure}{r}{0.2\textwidth}
\vspace{-20pt}
\begin{center}
\includegraphics[width=0.2\textwidth]{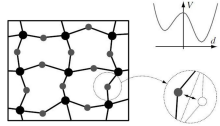}
\end{center}
\vspace{-20pt}
\caption{the potential in the Glass ,  P. W. Anderson, B. I. Halperin, and C. M. varma, \textit{Anomalous low-temperature thermal properties of glasses and spin glasses,Phil. Mag. } \textbf{25},1-9, 1972
}
\vspace{-10pt}
\end{wrapfigure}

in the solid two approximately equal metastable minimum around the the ideal binding axis has been created and the energetically lowest excitations of the system are transitions of individual atoms between nearly degenerate minima and the atoms flips around the binding axis.
in the other hand there is well know shape of potential which is similar to double well potential and it is the Higg's potential with Lagrangian 
$ L=\frac{1}{2}{\partial}_{\mu}\phi {\partial}^{\mu} \phi -\frac{1}{2}m^2 {\phi}^2- g^2 {\phi}^4 $ and there was some efforts in 70th to describe some concept with tunneling via this potential.

 We assume that the potential is even 
\begin{equation}
V(x)=V(-x)
\end{equation}
And if we denote the minimum of potential with $a$ and $-a$ and the second derivation of potential at these points with $V''(\pm a)={\omega}^2$we would like to estimate the quantity like 
\begin{equation}
\langle -a|e^{-\frac{HT}{\hbar}}| -a \rangle =\langle a|e^{-\frac{HT}{\hbar}}| a \rangle
\end{equation}
Or 
\begin{equation}
\langle -a|e^{-\frac{HT}{\hbar}}| a \rangle =\langle a|e^{-\frac{HT}{\hbar}}| -a \rangle
\end{equation}

By approximating the functional integral by it's semi classical limit, we need to at first step to calculate the classical Euclidean equation of motion, consistent with our boundary conditions.more precisely, there are three different types of classical solutions that fulfill the condition to be at coordinates $\pm a$ at times 0 ot $T$ one of them is where the particle rest at on top of one of the hill, $a$ the second one corresponding the similar situation with $-a$. But this two trivial solutions which are that the particle stay on the top of the hills are not our interest we can see they add zero amount to classical action.however there is another potentially interesting solution, where the particle start it’s motion at the top of one hill at $t=-T/2$ and reach to the other top on time $t'=T/2$. Then we limit $T$ to infinity.
\begin{center}
\begin{figure}[h]
\includegraphics[height=45mm]{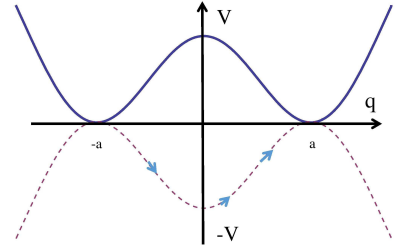}
\caption{Motion of Particle in the inversed Potential in $[-T/2, T/2]$}
\end{figure}
\end{center}

By the discussion in Section \ref{sec3}  we should set the energy to zero, thus the equation of motion takes the simple form 
\begin{equation}
\frac{dq}{dt}=(2V)^{\frac{1}{2}}
\end{equation} 
Or equivalently 
\begin{equation}
t-t_0=\int_{0}^{q} dq'(2V)^{-\frac{1}{2}}
\end{equation}
And $t_0$ is the constant of the integral.if we draw the $q$ verses the time, this takes the general form like the Fig.4 albeit it depend on the exact shape of the potential but the while case have the general feature which near the the minimum points of potential we have 
\begin{equation}
a-q \sim e^{-\omega t}
\end{equation}
which as usual $V''(\pm a)={\omega}^2$. It seems they are well localized particles in scale of $1/{\omega}$ and the tunneling takes place on this time scale.
%



\begin{center}
\begin{figure}[h]
\includegraphics[height=45mm]{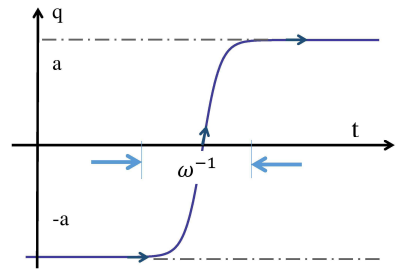}
\caption{Single Instanton }
\end{figure}
\end{center}
%
%

The solution which sketched in Fig.3 and 4  is called ``instanton" the name of instanton was invented by 't Hooft . The idea is that these objects are very similar in their mathematic structure  what are called solition , particle-like solutions of classical filed theories but unlike the soliton they are structure in time and the "instant-" come from it. Polyakov suggest this structure at first \citep{19}-\citep{1}-\citep{12}-\citep{14}-\citep{18}.

Simply  by replacing the $t$ to $-t$ we can have anti-instanton. It is easy to obtain the classical action as we discussed before 
\begin{equation}
S_0=\int dt[\frac{1}{2}(\frac{dq}{dt})^{\frac{1}{2}}+V]=\int dt (\frac{dq}{dt})^2=\int_{-a}^{a} dq (2V)^{\frac{1}{2}}
\end{equation}
We can see here that the action for two trivial solution of classical path are zero because $\dot{q}=0$. Also notices that $S_0$ is determined by the functional profile of the potentiall $V$ and it doesn't depend on the structure of the classical solution.

The confinement of the insatnton configuration to a narrow interval of time has an important implication. This is the critical importance because it means that for large $T$, the instanton and anti-Instanton are not the only approximate solutions of the equation of motion, they are also approximate solutions consisting of strings of sidely separated instanton and anti-instanton. We should summing over all such configurations with n objects centered at $t_1,t_2,\cdots ,t_n$ where
\begin{equation}
-T/2<t_n< \cdots <t_2<t_1<T/2
\end{equation}

For $n$ widely separated objects $S$ is $nS_0$  to evaluating the determinant we consider it has the form 
\begin{equation}
(\frac{\omega}{\pi \hbar})^{\frac{1}{2}}e^{-\frac{\omega T}{2}}K^n
\end{equation}
we evaluate $K$ later \citep{1} . This ansaltz has the simple reason. If we consider the potential well which has the shape similar to even function like the quadratic form the we expect that the eigenvalue of energy be similar to harmonic oscillator eigenvalue which are $\hbar \omega$ . To approve it if we following the path integral formalism it is easy to verify that for the a bit general form of well potential $V$ with $V''(0)={\omega}^2$ we have 
\begin{equation}\label{sss}
\langle 0| e^{-\frac{HT}{\hbar}}|0\rangle =N[det(-{\partial}_t^2+{\omega}^2)]^{-\frac{1}{2}}[1+O(\hbar)]
\end{equation}
But we know from the harmonic oscillator propagator that 
\begin{equation}
N[det(-{\partial}_t^2+{\omega}^2)]^{-\frac{1}{2}}=(\frac{m \omega}{\pi \hbar})^{\frac{1}{2}}e^{-\frac{\omega T}{2}}
\end{equation}
Now, if we consider that the proper Hamiltonian has the Eigensate of energies $|n\rangle$ with eigenvalues $n$ thus 
\begin{equation}
H|n\rangle =E_n |n\rangle 
\end{equation} 

Then 
\begin{equation}
\langle q'|e^{-\frac{HT}{\hbar}}|q \rangle =\sum_{n} e^{-\frac{E_n T}{\hbar}}\langle q'|n\rangle \langle n | q \rangle 
\end{equation}

for large $T$ the leading term belong to the least Eigenvalue, $E_0$ and therefore $\langle q'|e^{-\frac{HT}{\hbar}}|q \rangle \sim e^{-\frac{E_0 T}{\hbar}} \langle q'|0\rangle \langle0  | q \rangle $ .comparing this result with Eq. \ref{sss} leads to 
\begin{equation}
E_0=\frac{1}{2}\hbar \omega [1+O(\hbar)]
\end{equation}
And 
\begin{equation}
|\langle x=0|n=0\rangle |^2=(\frac{\omega}{\pi \hbar})^{\frac{1}{2}}[1+O(\hbar)]
\end{equation}
These calculation approve our ansaltz.

Now in order to calculating the propagator 
 also we should notice we must integrate over the locations of the centers
\begin{equation}
\int_{-T/2}^{T/2}dt_1 \int_{-T/2}^{t_1}dt_2 \cdots \int_{T/2}^{t_n-1} dt_n =\frac{T^n}{n!}
\end{equation}

Also we are not free to distribute instantons and anti-instantons arbitrary. For example of we start out at $-a$ the first object we encounter must be an instanton the next one must be an anti-instanton and etc. If we add up all case we have 
\begin{equation}
\langle -a | e^{-\frac{HT}{\hbar}}|-a \rangle =(\frac{\omega}{\pi \hbar})^{\frac{1}{2}} e^{-\frac{\omega T}{2}} \sum_{even\, n } \frac{(K e^{-\frac{S_0}{\hbar}}T)^n}{n!}[1+O(\hbar)]
\end{equation}

while we have 
\begin{equation}
\langle \pm a | e^{-\frac{HT}{\hbar}}|-a \rangle =(\frac{\omega}{\pi \hbar})^{\frac{1}{2}} e^{-\frac{\omega T}{2}} \frac{1}{2}[e^{Ke^{-\frac{S_0}{\hbar}}T} \mp e^{-Ke^{-\frac{S_0}{\hbar}}T}]
\end{equation}

Therefore we see that we have two low-lying energy Eigenstates with energies 
\begin{equation}
E_{\pm}=\frac{1}{2}\hbar \omega \pm \hbar K e^{-\frac{S_0}{\hbar}}
\end{equation}

\subsection{ Calculating the K Coefficient }
 In order to  evaluating the K coefficient, we should study the differential equation $[-d_{t}^2 +V''(x_c)]x=\lambda x$ with single instanton, we called the classical answer $\bar{x}$ . Because of time-translation invariance, this equation necessarily possesseses an Eigen function of eigenvalue zero. It its easy to check that the following function satisfy this requirement 
\begin{equation}
x_1={S_0}^{-\frac{1}{2}}\frac{d\bar{x}}{dt}
\end{equation}

As it has the zero eigenvalue we should change the coordinate to omit this value.this action add Jacobian factor $\frac{S_0}{2\pi \hbar}$ to main formula. Now if we call the calculated determinant without zero eigenvalue by Prime notation then we have 
\begin{equation}
\langle a|e^{-\frac{HT}{\hbar}}|-a\rangle =NT(\frac{S_0}{2\pi \hbar})^{\frac{1}{2}}e^{-\frac{S_0}{\hbar}}(\text{det}[-{\partial}_{t}^2+V''(\bar{x})])^{-\frac{1}{2}}
\end{equation}
Therefor by comparing with what we defined we obtain 

\begin{equation}
K=(\frac{S_0}{2\pi \hbar})^{\frac{1}{2}}\sqrt{\frac{\text{det}(-{\partial}_{t}^2+{\omega}^2)}{\text{det'}(-{\partial}_{t}^2 +V''(\bar{x}))}}
\end{equation}


\subsection{Computing the Determinant}
Here we want to study the differentiall equation likes the form of 
\begin{equation}
[-{\partial}_{t}^2+W]\psi=\lambda \psi
\end{equation}

Where $W$ is a bounded function of $t$ .we call ${\psi}_{\lambda}$ as the solution of our equation while boundary conditions are 
\begin{equation}
{\psi}_{\lambda}(-\frac{T}{2})=0 , \ \ \ \ \ \ \ \frac{\partial {\psi}}{\partial t}(-\frac{T}{2})=1
\end{equation}
The operator $[-{\partial}_{t}^2+W]$ has an Eigenvalue , $\lambda_{n}$, if and only if 
\begin{equation}
{\psi}_{\lambda_n}(T/2)=0
\end{equation}

Now, if $W^{(1)}$ and $W^{(2)}$ be two function of $t$ and the corresponding Eigenfunction of them be ${\psi}_{\lambda}^{1}$ and ${\psi}_{\lambda}^{2}$ the it is possible to proof that there is this the following statement \citep{1}
\begin{equation}
\text{det}[\frac{-{\partial}_{t}^2+W^{1}-\lambda}{-{\partial}_{t}^2+W^{2}-\lambda}]=\frac{{\psi}_{\lambda}^1(T/2)}{{\psi}_{\lambda}^2(T/2)}
\end{equation}
This valuable theorem help us to continue our complicated calculation. Know we want to estimate the determinant which the zero mode of it has been omitted, for this reason we estimate the determinant in the finite range of time $[T/2,-T/2]$ and divide it by it's smallest eigenvalue, ${\lambda}_{0}$ and then letting to $T$ to go to infinity . Thus we should construct solution of $[-{\partial}_{t}^2+U''(x_c)]\psi=\lambda \psi$ where $x_c$ is the classical solution.we can estimate the solution of $\lambda=0$ , which is the classical solution. By defining the $S_0$ 
\begin{equation}
S_0=\int dt[\frac{1}{2}(\frac{dx}{dt})^2+V]=\int dt (\frac{dx}{dt})^2=\int dx\sqrt{2V }
\end{equation}

One of the solution is 
\begin{equation}
x_1=S_{0}^{-\frac{1}{2}}\frac{d x_c}{dt} \rightarrow Ae^{-|t|}, t\rightarrow \pm \infty
\end{equation}

Here the $A$ is constant. As we have second order differential equation then we should find another solution for it . we set the second solution which the value of Wronskian given by
\begin{equation}
x_1{\partial}_{t} y_1-y_1 {\partial}_t x_1=2A^2
\end{equation}
Which yields to 
\begin{equation}
y_1 \rightarrow \pm Ae^{|t|}\, , t \rightarrow \pm \infty
\end{equation}
 To find the lowest eigenvalue we must find ${\psi}_{\lambda}(t) $ for small $\lambda$ .this can be done by standard method. We turn the differential equation to integral equation and iterate it once 
\begin{equation}
{\psi}_{\lambda}={\psi}_{0}(t)- {\lambda}(2A^2)^{-1} \int_{-T/2}^{t} dt'[y_1(t)x_1(t')-x_1(t)y_{1}(t')]{\psi}_{0}(t')
\end{equation}

Plus terms of the order ${\lambda}^2$ which we neglect. Thus 
\begin{equation}
{\psi}_{\lambda}(T/2)=1-{\lambda}(4A^2)^{-1} \int_{-T/2}^{T/2} dt[e^{T}x_{1}^2-e^{-T}y_{1}^2]
\end{equation}

For large $T$ the second term in integral can be neglected, thus 
\begin{equation}
{\psi}_{\lambda}(T/2)=1-\lambda (4A^2)^{-1}e^T
\end{equation}
Thus the lowest eigenvalue is given by 
\begin{equation}
{\lambda}_{0}=(4A^2)e^{-T}
\end{equation}

Therefor for large $T$ the modified determinant is given by 
\begin{equation}
\frac{det[-{\partial}_{t}^2+U''(x_c)]}{det[-{\partial}_{t}^2+{\omega}^2]}=\frac{{\psi}_{0}(T/2)}{\frac{{\lambda}_0e^{T}}{2}}=\frac{1}{2A^2}
\end{equation}

\subsection{Unstable States and Bounces}

Now we are ready to study the potential likes one has been sketched in Fig.5

\begin{center}
\begin{figure}[h]\label{YaZahra08}
  \includegraphics[height=45mm]{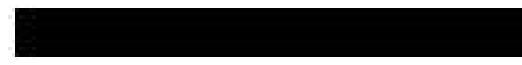}
  \caption{
 Special case of potential which provide the bounces motions, it has local minima but not absolute minimum
 }
  \end{figure}
\end{center}


we want understand instanton model by this kind of potential.if we neglect to the barrier penetration and we consider that the barrie has infinit height then we conclude the low eigenvalue of energy located at the minimum of potential and if we consider the real shape of potential then we should add correction to prevoius terms.to understand the classical solution we need to consider inverse of potential 


such the way we did for single instanton we should applye here and one maybe obtain 
\begin{equation}
\langle 0|e^{-\frac{HT}{\hbar}}|0\rangle =(\frac{\omega}{\pi \hbar})^{\frac{1}{2}}e^{-\frac{\omega T}{2}}e^{KTe^{-\frac{S_0}{\hbar}}}
\end{equation}

And thus one may deduce form result of Instanton model that the ground state of energy is similarly is
\begin{equation}\label{lll}
E_0=\frac{1}{2}\omega \hbar +\hbar K e^{-\frac{S_0}{\hbar}}
\end{equation}

 But there is some error in this conclusion. In the next section we explain why we need to correct our result.

%
%

\section{Decay Rate }

In obtaining the Eq.\ref{lll} it seems there are some mistake \citep{1}, at first the second term which we have calculated is small compared to the term $O({\hbar}^2)$ and it seem we must neglect it, this hint is similar to the section ??discussion on accuracy which the second term just gets meaning when we want to estimate the difference of two level of energy, $\Delta E$ .also in the bounce motion despite the previous instanton model in the turning point we have a node therefor it seems this motion is not Eigen function of lowest eigenvalues that is to say there must be a negative eigenvalue.thus $K$ which is inversely related to square roots of the eigenvalue, is imaginary. And the most important fact is that the eigenvalue we attempt to compute in nowhere to be found in the spectrum of Hamiltonian, because the states that we are studying is rendered unstable by barrier penetration.
thus it is only to be expected that $K$ should be imaginary. in the other hand it is true that the calculated term is small compared to real part of $E_0$ but it is leading contribution to the imaginary part of $E_0$ .therefor thecorrect version is 

\begin{equation}
\text{Im} E_0=\frac{\Gamma}{2}=\hbar |K| e^{-\frac{S_0}{\hbar}}
\end{equation}
Where $\Gamma$ is as usual the width of the unstable state. But it seems it is not consistent with reality and the correct for must be 

\begin{equation}
\Gamma=\hbar |K|e^{-\frac{S_0}{\hbar}}
\end{equation}
it seems the factor $\frac{1}{2}$ has been missed.to show the reason for this factor we need careful argument.

if we consider the following integral for real parameter $z$ 
\begin{equation}
J=\int dz (2\pi \hbar)^{-\frac{1}{2}} e^{-\frac{S(z)}{\hbar}}
\end{equation}

the $z$ descibe the path and $S(z)$ action of proper path. For getting intuition if we draw three sample of path in Fig.6

\begin{center}
\begin{figure}[h]\label{YaZahra006}
\includegraphics[height=35mm]{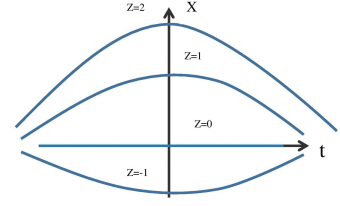}
\caption{Some different paths with various $z$ factor
}
\end{figure}
\end{center}


This path includes two important functions that occur in the real problem, $x(t)=0$ at $z=0$ and the bounce at $z=1$ .furthermore, the path is such that the tangent vector to the path at $z=1$ is $x_0$ .thus the path goes through the bounce in the most dangerous direction. That direction with which the negative eigenvalue is associated and $z=1$ is a maximum of $S$ .  $S$ goes to minus infinity as $z$ goes to infinity because the functions spend more and more time in the region beyond the turning point, where $V$ is negative.note that this implies that our integral divergent .if $x=0$ were the absolute minimum of $V$, that is to say, if $V$ were as shown in Fig.7
\begin{center}
\begin{figure}[h]
\includegraphics[height=45mm]{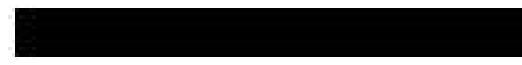}
\caption{The proper shape of potential which has absolute minimum }
\end{figure}
\end{center}

%
and there would be no divergence in integral.now let us suppose we analytically change $V$ in some way such that we go from this situation back to the one of interest.to keep this integral convergent we must distort the right-hand portion of the contour into the complex plane.if we assume that it is distorted into upper help plane. Following the standard procedure of method of steepest descents we led the contour along the real axis to $z=1$ , the saddle point, and the out along a line of constant imaginary part of $S$ .
The integral thus acquires an imaginary part, in the steepest descent approximation

\begin{eqnarray}
\text{Im} J &&=\text{Im} \int_{1}^{1+i\infty} dz(2\pi \hbar)^{-\frac{1}{2}}e^{-\frac{S(1)}{\hbar}}e^{-\frac{1}{2}S''(1)(z-1)^2/\hbar}\nonumber \\
&&= \frac{1}{2} e^{-\frac{S(1)}{\hbar}}|S''(1)|^{-\frac{1}{2}}
\end{eqnarray}

The factor $\frac{1}{2}$ arise because the integration is over only half of Gaussian peak.

In the stable situation, when the height of barrier penetration goes to infinity the solution of Schrödinger equation, corresponding to the ground state energy $E_0$ behaves as 
\begin{equation}
{\psi}_0(t) \sim e^{-\frac{iE_0t}{\hbar}}
\end{equation}
But for the case that we have not absolute minimum, $E_0$ becomes imaginary. Therefore for long times we have 
\begin{equation}
|{\psi}_0(t)|\sim e^{-\frac{Im E_0 t}{\hbar}}
\end{equation}
It clearly shows that the amplitude and therefore the probability of state decays. The parameter $|\frac{\hbar}{\text{Im}E_0}|$ is the lifetime of a now metastable state with wave function $\psi(t)$ . Let us to point out that the decay of state receives contributions from the continuation of all excited states. However, one expects, for intuitive reasons, that when the real part of the energy increases the corresponding contribution decreased faster with time, a property that can, indeed, be verified in examples. Thus, for large times, only the component corresponding to the pseudo-ground state survives. 

Also, according to the path integral scheme, the survival probability, defined by probability amplitude of remaining at the potential minimum $q_m$ which mean we should estimate the propagator $U(q_m,q_m,t)$ . by recent discussion one can obtains the survival probability 
\begin{equation}
U(q_m,q_m, \tau) \sim e^{-\frac{i\omega t}{2}}e^{\tau |K| e^{-\frac{S_0}{\hbar}}}
\end{equation}

Which if we return to real time, we have 
\begin{equation}
U(q_m,q_m, \tau) =(\frac{\omega}{\pi \hbar})^{\frac{1}{2}} e^{-\frac{i\omega t}{2}}e^{-\Gamma t}
\end{equation}
Where $\Gamma$ can given by 
\begin{equation}
\Gamma=\hbar |K|e^{-\frac{S_0}{\hbar}}
\end{equation}


\section{Discussion on Acuracy}

1.In practice   we have not good reason to keep the second terms. Because the exponential decays rapidly more than each function such as $\hbar$ also we should notice that this term is much smaller than $O(\hbar)$ which the correction for the first term .However it is the leading contribution to the difference of the energies , $\Delta E=E_+-E_- =\hbar K e^{-\frac{S_0}{\hbar}} $.

2.Our calculation agrees with the result from WKB-tyoe analysis of the tunneling process.in Appendix C we have solved it by WKB method for getting more insight in this problem and studying the Eigenvalue of more than $E_0$ ,i.e. $E_n$ .but it may appear that these kind of calculation was abit overkill for describing the simple tunneling process. We have obtained result in Appaendix C by WKB method by more elementary means, why then did we discuss instantons at such length? in fact WKB is uncontrolled approximation in general and it is hard to say that the result of this methods is accurate or not .in the contrast the model of instanton and approximation in path integral are controlled by number of well defined expansion parameters. For example we can go beyond the semi classical approximation and calculations in principle can drive to arbitrary accuracy.

3.Our approximation has been based on the assumption that the instanton and anti-instanton are all widely separated.we should verify that the major portion of our final result comes from configurations where this is indeed the case .for fixed value of $x$ in the exponential series $\sum \frac{x^n}{n!}$ the terms less than $n$ in order of $x$ have the main contribution and the next terms decrease rapidly .therfore in our claculation the terms are $n \lesssim KT e^{-\frac{S_0}{\hbar}}$ important. therefor the density of instantons, $\frac{n}{T}$ are small and our approximation is valid.

4.in our study, we want to apply this method to our tilted-washboard case in Josephson-junction which related to superconductivity at low temperature. If we develop the approximation to claculate the value of the state of energy with accuracy of order $O(\hbar)$ does it correct action? in fact we have the thermal noise in the system if we want to calculate the difference of energy with accuracy $\Delta E$ we have to provide the situation which 

\begin{equation}
\Delta E \gg {\Delta} E_{\text{Termal}}=kT 
\end{equation} 

if we consider energy as $E_0={\epsilon}_0 +{\epsilon}_1 \hbar +{\epsilon}_2 {\hbar}^2 $ then if our calculation  want to be meaningful we need to have a situation which 
\begin{equation}
{\epsilon}_2 {\hbar}^2 \gtrsim KT 
\end{equation}
In typical mili-Kelvin temperature, it means that ${\epsilon}_2 \gtrsim 10^{42} $ which is impossible. the typical qubit energy is $E_{01}\sim 10GHz \sim 0.5K$ and the typical experimental temperature is $T\sim 0.02K$ .

5.In our calucaltion for phase Qubit we neglect from the resistance of the junction. but it is important to emphasize that we can do it for low temperature and only for slow time evolution of phase. therfor we should have $T, \omega  \ll  \Delta$.

\section{ Phase Qubit}

\subsection{Single Josephson-Junction Phase Qubits}

Single Josephson junctions phase Qubits consists of one Josephson junction which use the quantum tunneling effect to produce the continuous current in the existence of external current source, $I_e$.

\begin{center}
\begin{figure}[h]\label{YaZahra07}
\includegraphics[height=45mm]{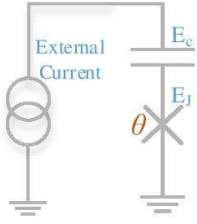}
\caption{The proper shape of potential which has absolute minimum }
\end{figure}
\end{center}
 As we will see at next subsection, we need to introduce the superconducting phase difference,or for simplicity phase,$\delta$ ( it sometimes written by $\theta$ or$\phi$) between two end point if junction which given by
\begin{equation}
\delta(t)=\frac{2e}{\hbar}\int V(t)dt + {\delta}_0
\end{equation}
Where the $V(t)$ is the voltage difference of junction.
As we will see by reasonable approximation the current of Josephson junction has the sinusoidal from 
\begin{equation}
I_J=I_c \sin{\delta}
\end{equation}
we will later disuse about correction term which this relation requires. $I_C$ is called critical current 
Therefore the leading equation in the circuit is 
\begin{equation}
\frac{\hbar }{2e}C \ddot{\delta}+\frac{{\hbar}^2}{2eR} \dot{\delta}+I_C \sin{\delta}=I_e
\end{equation} 
Where $I_e$ is the bias current. Also from this equation it is clear that the 
The potential energy corresponds to energy of the Josephson current can be written as
\begin{equation}
U(\phi)=E_J(1-\cos{\delta})-\frac{\hbar}{2e}I_e{\delta}
\end{equation}
Where $E_J=\hbar/2eI_e$ is the Josephson energy. This potential energy has a form of tilted-washboard, Fig.9 .
\begin{center}
\begin{figure}[h]
\includegraphics[height=70mm]{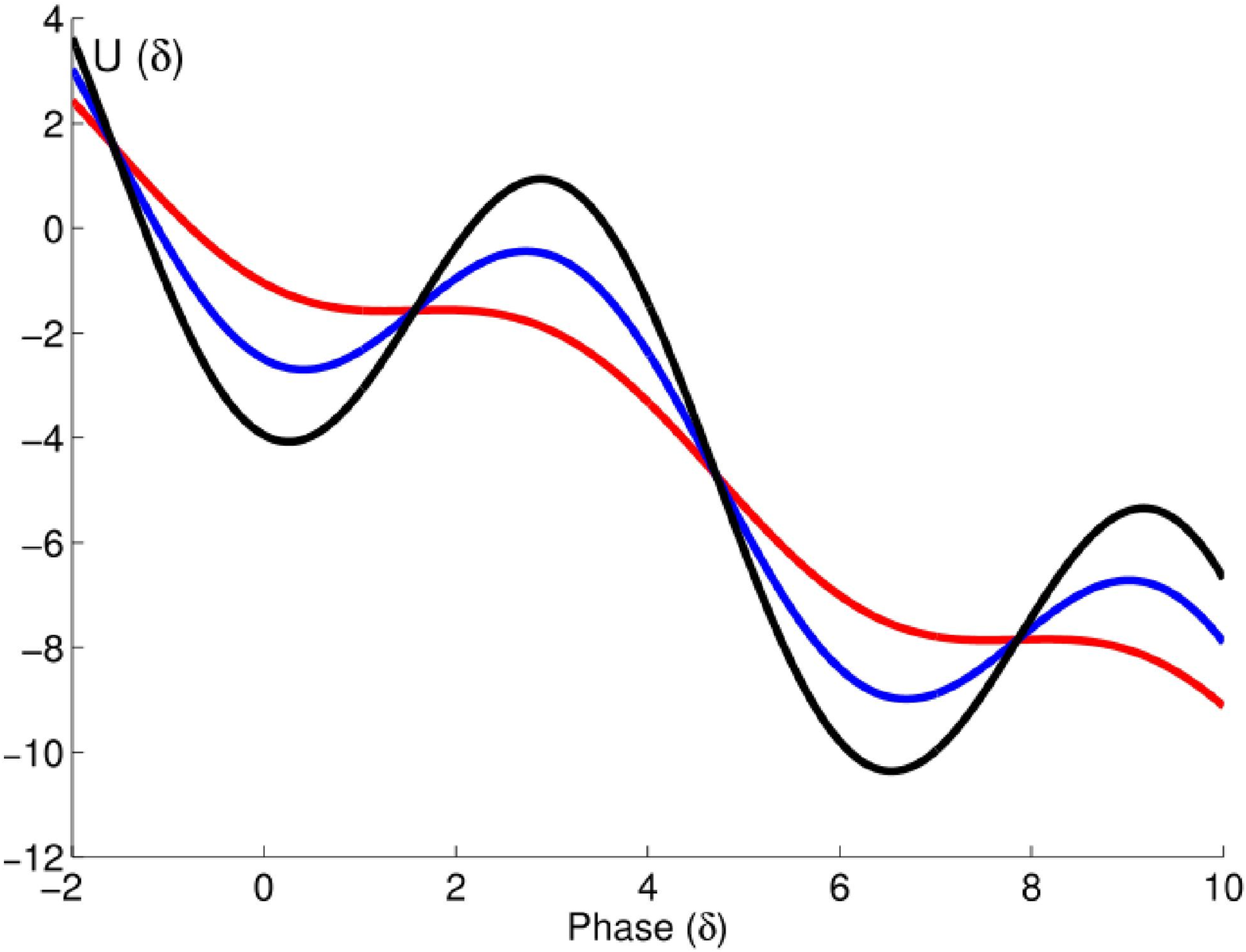}
\caption{Tilted-Washboard phase potential for various $I_e$ and $I_c$ . }{}
\end{figure}
\end{center}
The ratio of $\frac{I_0}{I_{dc}}$ determine the shape of the tilted-washboard potential. Our potential has the local minima, if in the relation $f(x):=-\alpha\cos{x}-\beta x $ , it must $f'(x):\frac{\beta}{\alpha} <1 $ and also in has the hill-shape if $f''(x)>0$ in some region of domain, which requires to $\alpha>0$ .


The Lagrangian corresponds to the differential equation of circuit can given by 
\begin{equation}
L=\frac{{\hbar}^2 {\dot{\delta}}^2}{4E_C}-E_j(1-\cos{\delta})+\frac{\hbar}{2e}I_e \delta
\end{equation}

And similarly the Hamiltonian of system can be written as 
\begin{equation}
H_{dc}=-E_C \frac{{\partial}^2}{\partial {\delta}^2}+E_J\cos{\delta}+ \frac{\hbar}{2e}I_e \delta
\end{equation}
In order to manipulate the system we need to evolve the system by time-dependent current, this , manipulation introduce the Hamiltonian of interaction which given by 
\begin{equation}
H_{\mu\nu}=\frac{{\Phi}_0}{2\pi} I_{\mu\nu} \delta=\frac{{\Phi}_0}{2\pi} I(t) \cos(\omega t +\phi ) \delta
\end{equation}
Here ${\Phi}_0=\frac{\hbar}{2e}$ is qunat of flux,Thus the total Hamiltonian of system yields 
\begin{equation}
H(t)=H_0+V(t)=H_0+ \frac{I_0 {\Phi}_0 }{2\pi} I(t) \cos(\omega t +\phi)
\end{equation}

%
%
%
%

\subsection{Correction Term  to  Josephson Junction Current }
In order to building the phase Qubits we use from Josephson junction. This junction play the fundamental rules in our analyses and construct our Hamiltonian, therefor it's important to study it more carefully. As we discussed in the subsection A there is a Sinusoidal current respect of phase, $J=J_C\sin(\phi) $.scrabble analysis can obtain it by considering the Schrödinger for the left and right hand of the junction as 

\begin{equation*}
i\hbar \frac{\partial {\psi}_1}{\partial t}=U_1{\psi}_1 +K{\psi}_2\\
\end{equation*}
\begin{equation}
i\hbar \frac{\partial {\psi}_2}{\partial t}=U_2{\psi}_2+K{\psi}_1
\end{equation}

If we apply the voltage between to pair of the junction then there must be different energy, $U_2-U_1=eV$ between them.by rewriting the equations we have 
\begin{equation*}
i\hbar \frac{\partial {\psi}_1}{\partial t}=-\frac{eV}{2}{\psi}_1 +K{\psi}_2\\
\end{equation*}
\begin{equation}
i\hbar \frac{\partial {\psi}_2}{\partial t}=\frac{eV}{2}{\psi}_2+K{\psi}_1
\end{equation}

one answer for this equation can be written as ${\Psi}_1=\sqrt{n_1} e^{i{\theta}_1}$ and ${\Psi}_2=\sqrt{n_2} e^{i{\theta}_2}$.by  substituting them in Schrödinger defining the phase as $\phi={\theta}_2-{\theta}_1$ we have 

\begin{equation*}
\frac{\partial n_1}{\partial t}=\frac{2}{\hbar}K\sqrt{n_1 n_2}\sin(\phi)
\end{equation*}
\begin{equation}
\frac{\partial n_2}{\partial t}=-\frac{2}{\hbar}K\sqrt{n_1 n_2}\sin(\phi)
\end{equation}
therfor the current in the josephson can be written as 
\begin{equation}
J=J_C \sin(\phi)
\end{equation}
which as called AC Josephson effect.by considering the $n_1=n_2$ we can observe the DC Josephson effect by finding the derivation of phase respect to time which yields 

\begin{equation}
\frac{\partial \phi}{\partial t}=\frac{2e}{\hbar}V
\end{equation}
Therefore the must general form of current would be 

\begin{equation}
J=J_C \sin({\phi}_0+\frac{2eV}{\hbar}t)
\end{equation}
this relation is famous be it is important to notice this relation is not exact description of the system and there is some correction to this relation.in order to obtaining the correction function to current respect to phase here we use from Ginzburg–Landau theory which it was postulated as a phenomenological model which could describe some kind of superconductors without examining their microscopic properties by Vitaly Lazarevich Ginzburg and Lev Landau.this theory later derived from BCS theory.
this model suggest the free energy for superconducting system with wavefunction $\psi$ as 

\begin{equation}
F=F_n+\alpha |\psi|^2+\frac{\beta}{2}|\psi |^4+\frac{1}{2m}|(-i\hbar \nabla -2e \mathbf{A})\psi |^2+\frac{|\mathbf{B}|^2}{2{\mu}_0}
\end{equation}
where $F_n$ is the free energy in the normal phase, $\alpha$ and $\beta$ in the initial argument were treated as phenomenological parameters, $m$ is an effective mass, $e$ is the charge of an electron.by varying the equation respect to $\psi$ or equivalently using Euler-Lagrange equation we will have 
\begin{equation}
\alpha \psi +\beta |\psi |^2 \psi +\frac{1}{2m}(-i \hbar \nabla -2 e \mathbf{A})^2 \psi =0
\end{equation}
and 
\begin{equation}
\mathbf{J}=\frac{2e}{m}\mathbf{Re}{{\psi} ^{*}(-i \hbar \nabla -2 e \mathbf{A})\psi }
\end{equation}
in the absence of electromagnetic fields previous differential equation can be written as
\begin{equation}
\alpha \psi +\beta |\psi |^2 \psi -\frac{{\hbar}^2}{2m}\frac{d^2 \psi}{d^2 x}=0
\end{equation}
if we demand that the wave function obey the usual boundary conditions in quantum mechanicsm which means vanishing itself and it's first derivation at infinity, then we would have 
\begin{equation}
\alpha {\psi}_{\infty} +\beta |{\psi}_{\infty} |^2 {\psi}_{\infty} =0
\end{equation}
and thus 
\begin{equation}
|{\psi}_{\infty} |^2=-\frac{\alpha}{\beta}
\end{equation}
if we rewrite Ginzberg-Landaue evolution equation by $f:=\frac{\psi}{{\psi}_{\infty}}$ and $\zeta:=\frac{\hbar}{\sqrt{2m|\alpha|}}$ we would have 
\begin{equation}
f-|f|^2f+{\zeta}^2 \frac{d^2 f}{d x^2}=0
\end{equation}
solving this differential equation analytically is impossible, in the easiest way if we assume that the width of junction,$L$, is more less that $\zeta$ then at first order of approximation we have 
\begin{equation}
\frac{d^2 f}{dx^2}=0
\end{equation}
Therefor the differential equation has a simple solution 
\begin{equation}
f(x)=ax+b
\end{equation}
boundary condition $f(0)=\frac{\psi(0)}{{\psi}_{\infty}}=1$and $f(L)=e^{i\delta}$ lead to write $f$ as 
\begin{equation}
f(x)=\frac{e^{i\delta}-1}{L}x +1
\end{equation}
Thus current must be
\begin{equation}
\mathbf{J}=\frac{2e}{m}\mathbf{Re}|{\psi}_{\infty}|^2(\frac{e^{-i\delta} -1}{L}x+1)(-i\hbar \frac{e^{i\delta}-1}{L})
\end{equation}
\begin{equation}
\mathbf{J}=\frac{e\hbar }{m L}|{\psi}_{\infty}|^2 \sin(\delta)
\end{equation}
it is the well known answer but as is clear it is not the exact answer.

we can better our approximation by neglecting just second term of differential equation
\begin{equation}
f(x)+{\zeta}^2 \frac{d^2 f(x) }{dx^2}=0=f(\frac{x}{\zeta})+\frac{d^2 f(\frac{x}{\zeta})}{d(\frac{x}{\zeta})^2}
\end{equation}
the answer in this regime is 
\begin{equation}
f(x)=\frac{e^{i\delta}-\cos(\frac{L}{\zeta})}{\sin(\frac{L}{\zeta})} \sin(\frac{x}{\zeta})+\cos(\frac{x}{\zeta}) \label{ali}
\end{equation}
and thus current must be 

\begin{equation}
\mathbf{J} =\frac{1}{\text{sinc}(\frac{L}{\zeta})}\frac{e\hbar}{m L}|{\psi}_{\infty}|^2 \sin(\delta)
\end{equation}
the sinusoidal shape of current has been kept here accidently.

in order to finding the correction function we must consider the whole terms in differential equation and solving the equation by iteration.the main idea is to add coefficient, $k$ , to second term of equation and rise it from $0$ to $1$ in $N$ steps.the answer of equation when $k=\frac{n}{N}=n\epsilon$ denotes by $f_n$ .thus the differential equation separate to two equation, imaginary and real part of $f$ 

\begin{eqnarray}
&&\delta f_r-\epsilon |f|^2 f_r +{\zeta}^2 \frac{d^2 \delta f_r}{dx^2}- \nonumber \\
&&n\epsilon (2|f|^2\delta f_r +\mathbf{Re}\{f\}^2 \delta {f_r} + \mathbf{Im}\{f^2\}\delta f_i)=0
\end{eqnarray}
و 
\begin{eqnarray}
&&\delta f_i-\epsilon |f|^2 f_i +{\zeta}^2 \frac{d^2 \delta f_i}{dx^2}-\nonumber \\
&&n\epsilon (2|f|^2\delta f_i -\mathbf{Re}\{f\}^2 \delta {f_i} + \mathbf{Im}\{f^2\}\delta f_r)=0
\end{eqnarray}
the correction function can be see in Fig 10 .it shows that we can have near $5\%$ error is some phases in our calculation if we consider current purely sinusoidal\citep{30} .
\begin{center}
\begin{figure}[h]
\includegraphics[height=61mm]{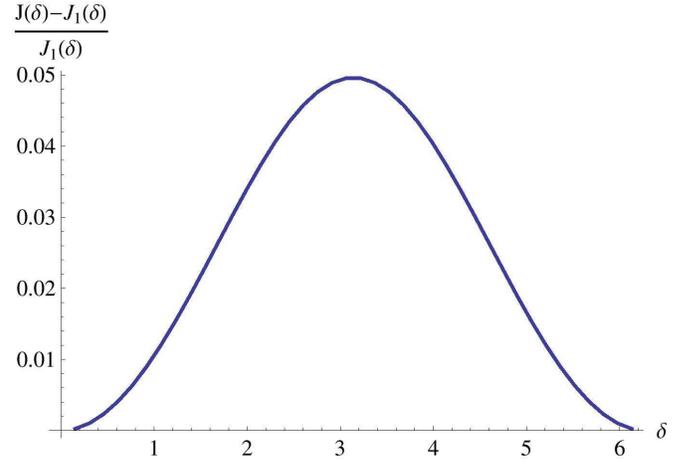}
\caption{Relative divergence from sinusoidal shape of current verses phase }
\end{figure}
\end{center}


\subsection{Tilted-Washboard Potential}

%
%
%
%
%
%
Now we know that the potential in Phase Qubit can given by  
\begin{equation}
V(t,q):=E_J\cos{q}+ \frac{I_{dc}{\Phi}_0}{2\pi}q +\frac{{\Phi}_0}{2\pi} I_{\mu\nu}(t) q+\epsilon(\delta)
\end{equation}
Where $\epsilon(\delta)$ shows the correction to sinusoidal current term in tilted-washboard potential.


If we tune the parameter, Our system can settle in one of the well of the tilted-washboard potential of phase Qubit and it is similar to potential well in quantum mechanics. From the quantum mechanic we know that is some condition our energy levels are discrete, this condition are provided in this case and for some level near to ground we can have discrete energy level, as we can see in Fig 11 . 
\begin{center}
\begin{figure}[h]
\includegraphics[height=50mm]{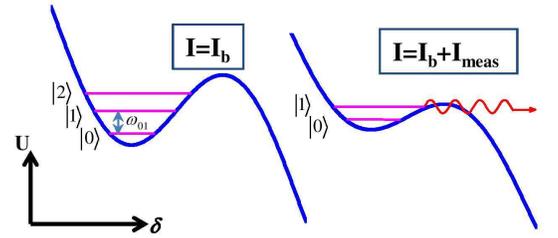}
\caption{Discreet levels of energy and the readout process}
\end{figure}
\end{center}

The properties of these levels and their wave-functions are our interest here. As we have learned because the shape of potential of Phase Qubits, we have bounce states here and we don't have stable states. Analysis of this structure has great value if we want to design precise Qubits and quantum gates. For applying the path integral, similar to approach we done in section, it is necessary to find the classical path of particle for this potential. The inversed potential has been drawn in Fig 12.
\begin{center}
\begin{figure}[h]
\includegraphics[height=40mm]{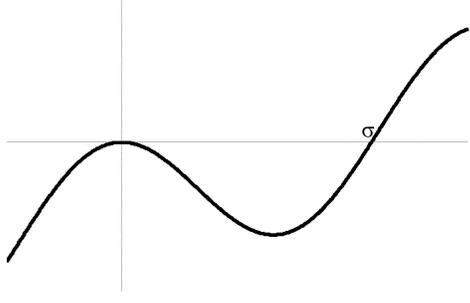}
\caption{Inversed potential of Phase Qubit}
\end{figure}
\end{center}
%
Here we have similar situation to unstable states and bounces, therefore we follow the mentioned solving way that we discussed in previous sections

In order to finding the classical path we should inverse the potential. If we call the turning point by $\sigma$, as is clear in fig 12 , then $\sigma$ is the zero of $V(x)=\alpha\cos{x}+\beta x+\epsilon(x)$ . The analytical solution of this equation obviously is not clear at first sight, specially the correction error, $\epsilon(x)$, has no simple formula. Thus it's better to solve it with soft wares, depend on our parameter. By knowing the turning point then estimating the action, $S_0$ is easy. as usual 
\begin{equation}
S_0=\int_{{\delta}_i}^{{\delta}_f} d\delta\sqrt{\frac{2V(\delta)}{m}}
\end{equation}
From our Hamiltonian it is celear that $m=\frac{{\hbar}^2}{2E_C}$ and $V(\delta)=E_J\cos{\delta}+\frac{\hbar}{2e}I_e \delta+\epsilon(\delta)$. Thus
\begin{equation}
S_0=\int_{0}^{\sigma} d\delta \sqrt{\frac{4E_C}{{\hbar}^2}}\sqrt{E_j\cos{\delta}+\frac{\hbar}{2e}I_e \delta + \epsilon(\delta)+c_0}
\end{equation}
Where $c_0$ is constant that appear form changing the coordinate in order to the hill point of potential locate at zero point of coordinate. Value of this integral can be calculate easily by soft wares.
Now we try to find the classical path. For simplicity and consistency we previous formula in previous subsection we change the variable of motion $x$ or $q$ instead of $\delta$ . From equation of motion we have 
\begin{equation}
\frac{1}{2}m{\dot{x_c}}^2=V(x_c)+E_0
\end{equation}
Thus the classical path obey from this relation
\begin{eqnarray}
t&&=t_1+\sqrt{\frac{2}{m}}\int_{0}^{x'} dx_c \sqrt{V(x_c)+E_0}\\
&&=\int_{0}^{x'} dx_c \frac{\sqrt{\frac{{\hbar}^2}{4E_C}}}{\sqrt{E_J \cos{x_c}+\frac{\hbar}{2e}I_e x_c +\epsilon(x_c)+E_0}}\nonumber 
\end{eqnarray}

The $E_0$ is the constant of motion in must be selected which in $x\rightarrow 0$, $t\rightarrow -\infty$ and vice-versa.
As we know the zero Eigenfunction of $[-{\partial}_t^2+V''(x_c)]$ is 
\begin{equation}
x_1={S_0}^{-\frac{1}{2}}\frac{dx_c}{dt}
\end{equation}
For the next estimation we strongly to know the behavior of $x_1$ respect to time. from previous equation we have function $t(x_c)$ and what we need is the inverse of this function $ g=f^{-1}=x_c(t)$. Finding the analytical form for this function is complicated and it's better to solve it numericaly in exact case that we need and with desire parameters. the total shape this function is like as Fig .13
\begin{center}
\begin{figure}[h]
\includegraphics[height=60mm]{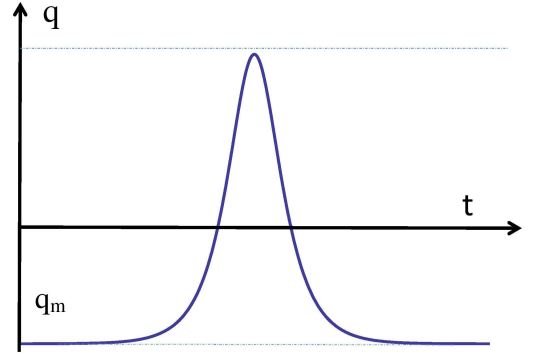}
\caption{Bounce motion}
\end{figure}
\end{center}

%
And for example, for the potentional $V(x)=\frac{1}{2}x^2+\frac{1}{2}g x^4$ the $x_c(t)$ has the form $x_c(t)=g(t) \sim \frac{1}{\cosh(t-t_0)}$, Fig 13.
Hence we expect that $x_1$ behave exponentially when time goes to infinity.
\begin{equation}
x_1=S_{0}^{-\frac{1}{2}}\frac{d x_c}{dt} \rightarrow Ae^{-|t|}, t\rightarrow \pm \infty
\end{equation}
We consider estimating the quantities $S_0$ and $x_c(t)$ let us to estimate the $A$ factor which is constant and fundamentally is function of just $I_C$ and $I_e$ and capacitance and cross section area of Josephson junction.

As we discussed before 
\begin{equation}
\frac{det[-{\partial}_{t}^2+U''(x_c)]}{det[-{\partial}_{t}^2+{\omega}^2]}=\frac{1}{2A^2}
\end{equation}

\begin{equation}
K={(\frac{S_0}{2\pi \hbar})^{\frac{1}{2}}\sqrt{\frac{1}{2A^2}}}
\end{equation}
And thus, in conclusion
\begin{equation}
\Gamma =\hbar (\frac{S_0}{2\pi \hbar})^{\frac{1}{2}}\sqrt{\frac{1}{2A^2}}e^{-\frac{S_0}{\hbar}}
\end{equation}
Where, action, $S_0$ given by 
\begin{equation}
S_0=\int_{0}^{\sigma} d\delta \sqrt{\frac{4E_C}{{\hbar}^2}}\sqrt{E_j\cos{\delta}+\frac{\hbar}{2e}I_e \delta + \epsilon(\delta)+c_0}
\end{equation}
And constant coefficient,$A$ is 
\begin{equation}
A=\lim_{t\rightarrow \infty}\frac{S_{0}^{-\frac{1}{2}}\frac{d x_c}{dt}}{e^{-t}}
\end{equation}


\section{Charge Qubit}
A charge Qubit is formed by a tiny superconducting island in the simplest form and it has been coupled by a Josephson junction to a superconducting reservoir and via the capacitor we can tune the gate voltage.in the low temperature the spectrum of energy becomes more discrete, there is no possibility to tunneling the cooper pair into the island except in some certain gate voltages. Therefore the number of cooper-pair can be known precisely, usually the island without no cooper-pair called $|0\rangle$ state and the island which has one cooper-pair called $|1\rangle$.in the certain voltage these two state becomes degenerate and cooper-pair can tunnel continually therefore the state of island can be direct sum of $|0\rangle$ and $|1\rangle$. 

If we would like to a quantum description of the circuit of charge Qubit, as is presented in Fig.14 which two region of superconductor connected by Josephson junction whit critical current, $I_C$ and capacitance $C$. in the classical view the relation between voltage, charge and capacitance is $V=\frac{Q}{C}$.therefore the leading equation in circuit can be given by 
\begin{equation}
\frac{d^2 \phi }{d t^2}=\frac{2e}{\hbar C}I_C \sin{\phi}
\end{equation}

\begin{center}
\begin{figure}[h]
\includegraphics[height=45mm]{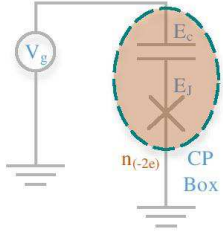}
\caption{Simple charge Qubit}
\end{figure}
\end{center}
One can suggest the  Lagrangian as 
\begin{equation}
L=\frac{1}{2}\frac{{\hbar}^2C}{4e^2}(\frac{d\phi}{dt})^2+\frac{\hbar I_C}{2e}\cos{\phi}
\end{equation}
It is also available to introduce the Hamiltonian of system. the canonical momentum, $\pi$ as usual define by 
\begin{equation}
\pi=\frac{\partial L}{\partial {\dot{\phi}}}=\frac{{\hbar}^2C}{4e^2}{\dot{\phi}}
\end{equation}
in the other hand
\begin{equation}
\frac{{\hbar}^2C}{4e^2}{\dot{\phi}}=\frac{\hbar CV}{2e}=\frac{\hbar q}{2e}=\hbar N
\end{equation}
Where $N$ in number of Cooper pairs.now by using the quantization rule $[\hat{\phi},\hat{\pi}]=i\hbar$ ,we convertt the classical Hamiltonian to quantum version of it,which yields
\begin{equation}
\hat{H}=E_C {\hat{N}}^2-E_J \cos{\hat{\phi}}
\end{equation}
here $\hat{\phi}$ and $\hat{N}=i \frac{d}{d\phi}$ .thus the Hamiltonian of charge Qubit in the simple case can be describe by 
\begin{equation}
\hat{H}=E_C(i\frac{d}{d\phi})^2-E_J \cos{\phi}
\end{equation}

path integral method can be apply to studying the charge Qubit easily.using the relation from Instanton model .

In the charge qubit with have periodic potential. Analysis of periodic potential is famous problem in Condensed-Mather filed and using the Bloch theorem help us to find the properties of the system. Here we try to find the spectrum of energy by method we have developed in the previous sections, Instanton model.if we ignore barrier penetration, the energy Eigen states are an infinitely degenerate set of states, each concentrates at the bottom of one of the wells. Barrier penetration changes this single eigenvalue into a continuous translations, the Bloch waves.the Instanton are much the same as in the preceding problem, the only novelty is that the Instantons can begin at any initial position $x=\lambda$ and go to the next one $x=\lambda+1$ like wise the anti-instantons can go from $x=\lambda$ to $x=\lambda-1$.otherwise,everything is as before. There is no constraint that instantons and anti-instantons must alternate. Of course, as we go along the line, each instanton or anti-instanton must begin where its predecessor ended. Furthermore, the total number of instantons minus the total number of anti-instantons must be equal the change in $x$ between the initial and final position Eigen states.thus we obtain 
\begin{eqnarray}
\langle {\lambda}_{+}|e^{-\frac{HT}{\hbar}}|{\lambda}_{-} \rangle &&= (\frac{\omega}{\pi \hbar})^{\frac{1}{2}}e^{-\frac{\omega T}{2}}
\sum_{n=0}^{n=\infty} \sum_{\bar{n}=0}^{\infty} \frac{1}{n!\bar{n}!} 
\nonumber 
\\
&& \times (Ke^{-\frac{S_0}{\hbar}})^{n+\bar{n}}{\delta}_{n-\bar{n}-{\lambda}_{+}+{\lambda}_{-}}
\end{eqnarray}
where $n$ is the number of instantons and $\bar{n}$ the number of anti-instantons. If we use the identity
\begin{equation}
{\delta}_{ab}=\frac{1}{2\pi}\int_{0}^{2\pi}d\theta e^{i\theta(a-b)}
\end{equation}
the sum becomes two independent exponential series, and we find

\begin{eqnarray}
\langle {\lambda}_{+}|e^{-\frac{HT}{\hbar}}|{\lambda}_{-} \rangle &&= (\frac{\omega}{\pi \hbar})^{\frac{1}{2}}e^{-\frac{\omega T}{2}}
\int_{0}^{2\pi} e^{i({\lambda}_{-}-{\lambda}_{+}}\frac{d\theta}{2\pi}
\nonumber 
\\
&& \times e^{2KT\cos{\theta} e^{-\frac{S_0}{\hbar}}}
\end{eqnarray}
thus we find a continuum of energy eigen states labeled by the angle $\theta$.the energy eigenvalues are given by 

\begin{equation}
E(\theta)=\frac{1}{2}\hbar \omega +2\hbar K \cos{\theta} e^{-\frac{S_0}{\hbar}}
\end{equation}
also the eigen function can be obtain by 
\begin{equation}
\langle \theta | \lambda \rangle =(\frac{\omega}{\pi \hbar})^{\frac{1}{2}}(2\pi)^{-\frac{1}{2}}e^{ij\theta}
\end{equation}


\section{Flux Qubits}
\begin{center}
\begin{figure}[h]
  \includegraphics[height=50mm]{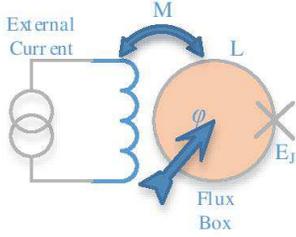}
  \caption{Simple Flux Qubit}
  \end{figure}
\end{center}



Flux Qubits are one of the main configuration in building the superconducting quantum circuits which represented in Fig 15. This circuit measure the flux that exist 
 in the circuit. the corresponding current in inductor with inductance ,$L$, is given by

\begin{equation}
I_L=\frac{\hbar}{2eL}(\phi -{\phi}_e)
\end{equation}
Where ${\phi}_e=\frac{2e}{\hbar}{\Phi}_e$ and ${\Phi}_e$ is the external magnetic flux. leading equation in the circuit thus given by 
\begin{equation}
\frac{\hbar}{2e}C{\ddot{\phi}}+\frac{\hbar}{2eR}{\dot{\phi}}+I_c\sin{\phi}+\frac{\hbar}{2eL}(\phi -{\phi}_e)=0
\end{equation}
Thus the  Hamiltonian of Flux Qubits given by 
\begin{equation}
H=\frac{{\hbar}^2}{4E_C}{\dot{\phi}}^2+E_J(1-\cos{\phi})+E_L\frac{(\phi-{\phi}_e)^2}{2}
\end{equation}
Fig 16 shows the potential of flux Qubit.  

\begin{center}
\begin{figure}[h]
\includegraphics[height=50mm]{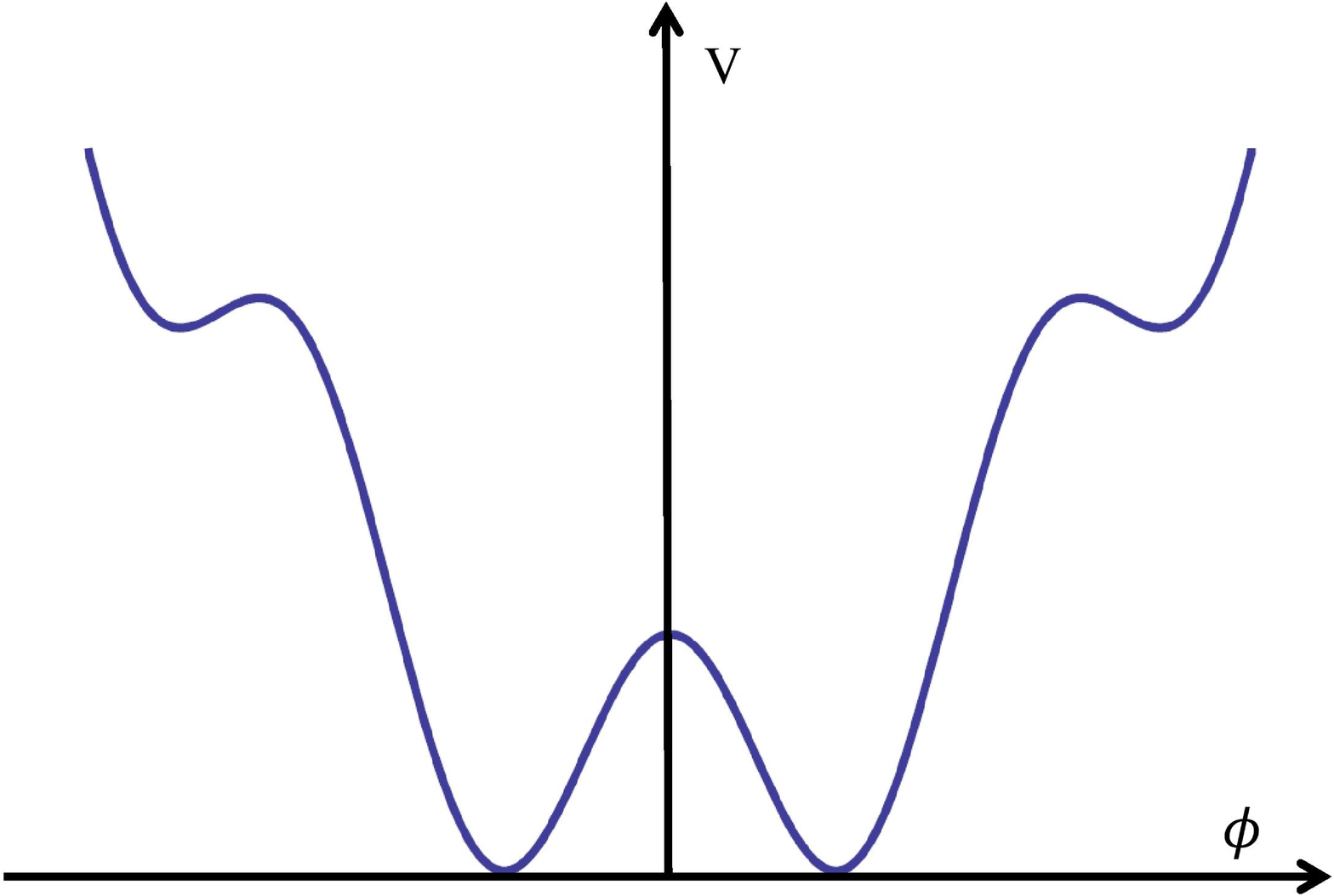}
\caption{Flux Qubits potential}
\end{figure}
\end{center}

Finding the spectrum of energies and wave-functions in this case is more routine, there are approximated method and almost reliable results for studying this kind of potential. Here the states are stable and we don't need more effort to obtain the Eigen energies.

\begin{eqnarray}
V(\phi)&&=E_J(1-\cos{\phi})+E_L \frac{({\phi}-{\phi}_e)^2}{2} \nonumber \\
&&\simeq E_J[\frac{{\phi}^2}{2!}-\frac{1}{4!}{\phi}^4]+E_L \frac{(\phi-{\phi}_e)^2}{2}
\end{eqnarray}

We have solved this problem in the Aappendix C with WKB method in details and here we discuss on a bit different and fast way by mixing the idea of WKB method and valuable result that we obtained in the path integral in previous sections.
We wish to study  the solution of 
\begin{equation}
-\frac{1}{2}{\hbar}^2 {\partial}_{x}^2 {\psi } +V{\psi} =E\psi
\end{equation}
Neat the bottom of each well there are two turning points and they are not separated with many wavelength, the consideration which we always keep it in the WKB method. Therefore it seems we cannot use the standard connection formulae for linear turning point. As alternative way we may safety approximate the potential by a harmonic oscillator potential. Thus for $x$ near to the bottom of well, $a$, we may write 
\begin{equation}\label{aaa1}
-\frac{1}{2}{\hbar}^2 {\partial}_{x}^2 {\psi } +\frac{1}{2}(x-a)^2{\psi} =E\psi
\end{equation}
Now we try to match the solution of WKB outside the well to the solution of equation \ref{aaa1}
\begin{equation}
{\psi}_{\pm}=k^{-\frac{1}{2}}[e^{{\hbar}^{-1}\int_{0}^{x} kdx' \pm e^{-{\hbar}^{-1} \int_{0}^{x} kdx'}}]
\end{equation}
We construct the even and odd WKB solution for the region $0 \leq x \leq a$. If we define $k(x)=\sqrt{2(V-E)}$ then
\begin{equation}
{\psi}_{\pm}={k}^{-\frac{1}{2}} [e^{\frac{1}{\hbar}\int_{0}^{x}kdx'}\pm e^{-\frac{1}{2}\int_{0}^{x} kdx'}]
\end{equation} 
We are interested in the solution which $E$ is of order $E$ therefore we can expand the integral to first order of $\hbar$. It lets to us to write 
$k=(2V)^{\frac{1}{2}}-E(2V)^{-\frac{1}{2}}$.
We separate the terms in exponential to $E$ dependent and independent. For independent term we have 
\begin{eqnarray}
\int_{0}^{x} dx (2V)^{\frac{1}{2}} &&=\int_{0}^{a} dx(2V)^{\frac{1}{2}} -\int_{a}^{x} dx (2V)^{\frac{1}{2}}
\\
&&=\frac{1}{2}S_0-\frac{1}{2}(a-x)^2
\end{eqnarray}
For the next term, one can use the equation  .Thus we obtain 
\begin{eqnarray}\label{aaa2}
{\psi}_{\pm}&&=(a-x)^{-\frac{1}{2}}[e^{{\hbar}^{-1}[\frac{1}{2}S_0-\frac{1}{2}(a-x)^2+E\ln {S_0}^{-\frac{1}{2}A^{-1}(a-x)}]}
\nonumber 
\\
&& \pm e^{-{\hbar}^{-1}[\frac{1}{2}S_0-\frac{1}{2}(a-x)^2+E\ln {S_0}^{-\frac{1}{2}A^{-1}(a-x)}]}]
\end{eqnarray}

if we write 
\begin{equation}
E=\hbar (\frac{1}{2}+\epsilon)
\end{equation}
We have considred $\omega=1$. Then the Equation \ref{aaa2} becomes
\begin{eqnarray}
{\psi}_{\pm} &&=\{e^{\frac{S_0}{\hbar}{S_0}^{-\frac{1}{4}}A^{-\frac{1}{2}}e^{-\frac{(a-x)^2}{2\hbar}}}
\nonumber 
\\
&& \pm (a-x)^{-1}e^{-\frac{S_0}{2\hbar}}{S_0}^{\frac{1}{4}}A^{\frac{1}{2}}e^{\frac{(a-x)^2}{2\hbar}}\}[1+O(\epsilon)]
\end{eqnarray}
Now in order to connect the solution of Eq.\ref{aaa1} we need to find the solution of it at first step. For $E=0$ we know it, $\psi_1=e^{-\frac{(a-x)^2}{2\hbar}}$ . Of course there is another solution ,${\phi}_1$, which is odd. This does not a simple form in terms of elementary functions. But it's asymptotic form 
\begin{equation}
{\phi}_1=(a-x)^{-1}e^{\frac{(a-x)^2}{2\hbar}}
\end{equation}
Note that we normalized it such that the Wronskian of the two solution becomes
\begin{equation}
{\phi}_1{\partial}_x {\psi}_1-{\psi}_1{\partial}_x{\phi}_1=\frac{2}{\hbar}
\end{equation}
We wish to solve the Eq.\ref{aaa1} By the same argument which we had in Section ??? 
\begin{equation}
{\psi}={\psi}_1-{\epsilon}\int_{x}^{\infty} dx'{\psi}_1(x')[{\psi}_1(x'){\phi}_1(x)-{\phi}_1(x'){\psi}_1(x)]
\end{equation}

We have chosen here the solution that vanish as $x$ goes to plus infinity. In this region we can use 
\begin{equation}
\int_{-\infty}^{\infty}dx {\psi}_1^2=\sqrt{\pi\hbar}
\end{equation}
To write 
\begin{eqnarray}
\psi =e^{-\frac{(a-x)^2}{2\hbar}}[1+O(\epsilon)]-\epsilon \sqrt{\pi \hbar}(a-x)^{-1} e^{\frac{(a-x)^2}{2\hbar}}
\end{eqnarray}
As it should be, this is proportional to Eq.\ref{aaa2} if we choose
\begin{equation}
E=\frac{1}{2}\hbar + \hbar e^{-\frac{S_0}{\hbar}}A(\frac{S_0}{\pi \hbar})^{\frac{1}{2}}
\end{equation}

\begin{center}
\begin{figure}[h]
\includegraphics[height=40mm]{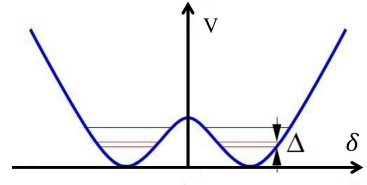}
\caption{Energy levels in flux Qubits}
\end{figure}
\end{center}

\section{Perturbation In Time}
If we want to build the quantum gate, what we should to do in practice is constructing the system that the corresponding Hamiltonian of system be similar to desired Hamiltonian in theory that we need. For this reason we can apply the interaction to system which the time evolution system at least behave what we expect from ideal time invariant gate. In the other hand the interaction Hamiltonian in fact palys the role of measurement. Therefore the Hamiltonian of the system is now time dependent. 
\begin{equation}
H(t,q)=H_0(q)+V(q,t)
\end{equation}

The simplest way to studying  the evolution of system, can be obtain by numerically integrating the corresponding time dependent Schrödinger 
Equation
\begin{equation}
i\hbar \frac{\partial }{\partial t}{\psi(x,t)}=[H_0(x)+v(x,t)]{\psi}(x,t)
\end{equation}
To compute the evolution of the population on the Eigen states of Qubits, the time dependent wave function is expanded in the Eigen states $|n\rangle$ of $H_0(q)$ 
\begin{equation}
{\psi}(q,t)=\sum_{n=0}^{N}C_{n}(t)|n \rangle 
\end{equation}
The expansion coefficient are obtained by solving the time dependent matrix equation in fact
\begin{equation}
i \frac{\partial}{\partial t}C_{n}(t)=\sum_{n'=0}^{N} H_{nn'}^RC_{n'}(t)
\end{equation}
Thus the matrix element of Hamiltonina  are our interest.
\begin{equation}
H_{nn'}^R(\tau)\sim [E_n {\delta}_{nn'}+\langle n|V(q,t)|n'\rangle ]
\end{equation}
specific quantum gates means specific Hamiltonian matrix elements, therefore we should control the interacetion potential,$V$, in order to get desire matrix element. it real problemm, especially in phas Qubits the state are time dependent too and if we want to be more careful it's better to use the path integral formalism, as we know 
\begin{equation}
\label{bbbbb}
\frac{\partial U(t,t')}{\partial t}|_{t=t'}=-\frac{H(t)}{\hbar}
\end{equation}
Then in order to estimating the property of $H$ we have 
\begin{equation}
\frac{\partial }{\partial t}\langle m |U(t,t')|n \rangle |_{t=t'}=-\frac{\langle m |H(t)|n \rangle }{\hbar}
\end{equation}

\section{Conclusion}

Now, at the end of this paper maybe merits of path integral has been more appear to us 
Whereas the classical limits is not always easy to retrieve within the canonical formulation of quantum mechanics, it constantly remains visible in the path integral approach. also the path integrals allow for an efficient formulation of non-perturbative approaches to the solution of quantum mechanics problems. We obtain the state and metastable state of Phase Qubits for the first time  and the properties of Flux and Charge which obtained with path integral method were consistent whit other method's result but with more visibility and more accuracy. in the other hand the path integral doesn't have time dependent or time independent case, it describe the evolution of system clearly without approximation which we introduce in usual quantum mechanic such as Dyson series and etc.  in the recent works we are trying to obtain the spectrum of Hamiltonian during the time to reveal how this method in valuable and how it can increase our accuracy as much as we need .

\begin{appendices}

\section{ Gaussian Integral}

The main idea of semi-classical approach in path integral is using the property of Gaussian integral and wick's theorem\citep{2.0}, we briefly discuss about it here.

The Gaussian integral 
\begin{equation}
Z(\mathbf A)=\int d^nx e^{-\sum_{i,j=1}^{n} \frac{1}{2}x_i{\mathbf A}_{ij} x_j}
\end{equation}
Converges if the matrix$ \mathbf{A}$ with elements $\mathbf{A}_{ij}$ is a symmetric complex matrix such
That the real part of the matrix is non-negative.this implies that all eigenvalues of $Re \mathbf{A}$ are non-negative and no eigenvalue $a_i$ of $\mathbf{A}$ vanishes

\begin{equation}
Re {\mathbf{A}} \geq 0 , \ \ \ \ \ \ \ \ a_i \neq 0
\end{equation}
In several way we can proof that 
\begin{equation}
Z(\mathbf{A})=(2\pi)^{\frac{n}{2}}(det \mathbf{A})^{-\frac{1}{2}}
\end{equation}

We start the proof with real matrix and in the next step, we develop it for imaginary matrix. Any real symmetric matrix can be diagonalized by an orthogonal transformation and the matrix $\mathbf{A}$ in can thus be written as 

\begin{equation}
\mathbf{A=ODO^{T}}
\end{equation}
where matrix $\mathbf{A}$ is orthogonal, $\mathbf{OO^T=I}$, and $\mathbf{D}$ is diagonal matrix with element : $\mathbf{D}_{ij}=a_i{\delta}_{ij}$.by changing the variable $\mathbf{x \mapsto y}$ in integral we have 

\begin{equation}
x_i=\sum_{j=1}^{n} {O}_{ij} y_{j} \rightarrow \sum_{i,j} x_i \mathbf{A}_{ij} x_j=\sum_{i,j} x_{i}O_{ik}a_{k} O_{jk} x_j=\sum_{i} a_{i} {y_i}^2
\end{equation}

The Jacobian of this transformation is 

\begin{equation}
J=|\text{det} O|=1
\end{equation}

Therefore 
\begin{equation}
Z(\mathbf{A})=\prod_{i=1}^{n} \int dy_{i} e^{-\frac{a_i {y_i}^2}{2}}
\end{equation}
As the matrix $\mathbf{A}$ is positive, all eigenvalues $a_i$ are positive thus the integral converge and we find that 
\begin{equation}
Z(\mathbf{A})=(2\pi )^{\frac{n}{2}}(a_1 a_2 \cdots a_n)^{-\frac{1}{2}}=(2\pi )^{\frac{n}{2}}(det \mathbf{A})^{-\frac{1}{2}}
\end{equation}

In the most general form we have 
\begin{equation}
Z(\mathbf{A,b})=\int d^n x e^{-\sum_{i,j=1}^{n} \frac{1}{2}x_i {\mathbf{A}_{ij} x_j +\sum_{i=1}^{n} b_i x_i}}
\end{equation}

To calculate this integral, one looks for the minimum of the quadratic form 
\begin{equation}
\frac{\partial}{\partial x_k}\bigg(\sum_{i,j=1}^{n} \frac{1}{2} x_i {\mathbf{A}}_{ij}x_j - \sum_{i=1}^{n} b_i x_i \bigg)=\sum_{j=1}^{n}{\mathbf{A}}_{kj} x_{j}-b_k=0
\end{equation}

Introducing the inverse matrix
\begin{equation}
\mathbf{\square}={\mathbf A}^{-1}
\end{equation}

one can write the solution as 
\begin{equation}
x_i=\sum_{j=1}^{n} {\mathbf \square}_{ij} b_j
\end{equation}

By changing the variable $x \mapsto y $ we have 
\begin{equation}
x_i=\sum_{j=1}^{n} {\square}_{ij} b_j +y_i
\end{equation}
Then the integral becomes 
\begin{equation}
Z(\mathbf{A,b})=(2\pi)^{\frac{n}{2}} (det \mathbf{A})^{-\frac{1}{2}} e^{\sum_{i,j=1}^{n}\frac{1}{2} b_i {\square}_{ij} b_j }
\end{equation}
In the case that matrix be imaginary the proof need more effort which available in Ref. \citep{2.0}

\section{ steepest desecnt method}

The steepest decent method is an approximation scheme to estimating certain types of contour integral in complex domain. It involve approximation integral by a sum of Gaussian expectation value.here we would like to analysis the real integral such as 
\begin{equation}
I(h)=\int_{a}^{b} dx e^{-\frac{A(x)}{h}}
\end{equation}
Which $A(x)$ is real function and we consider that $A(x)$ in analytic in the integral interval and it has absolute minimum in this region. We want to evaluate the integral in the limit of $h\rightarrow 0_{+}$ . in this limit the integral is dominated by the maximum of the integrand which it means minimum of $A(x)$ . if the minimum occur in $x_c$ then we have $A'(x_c)=0$ and generally $A''(x_c)>0$ .therefore the integration domain can restricted to a neighborhood $(x_c-\epsilon, x_c+\epsilon)$ which $\epsilon$ is arbitrarily small .as we can say 
\begin{equation}
A(x) \simeq A(x_c)+ \frac{1}{2}A''(x_c)(x-x_c)^2
\end{equation}
The the region that contributes is of order $\sqrt{h}$, thus it is convenient to change variable $x\mapsto y $ by $y=\frac{x-x_c}{\sqrt{h}}$.
Then the expansion converts to 
\begin{equation}
\frac{A}{h} =\frac{A(x_c)}{h}+\frac{1}{2}y^2 A''(x_c)+\frac{1}{6}\sqrt{h} A'''(x_c)y^3+ \frac{1}{24}A^{(4)}(x_c)y^4 +O({h}^{\frac{3}{2}})
\end{equation}
If we need to calculate just leading order therefore we have 
\begin{equation}
I(h) \simeq \sqrt{h} e^{-\frac{A(x_c)}{h}}\int_{\frac{-\epsilon}{\sqrt{h}}}^{\frac{\epsilon}{h}} dy e^{-\frac{A''(x_c)y^2}{2}}
\end{equation}

We can extend domain of integral to infinity because of for similar reason contributions from outside the integration domain are exponentially small in $\frac{1}{\sqrt{h}}$ .the leading contribution is thus given by Gaussian integral which yields
\begin{equation}
I(h) \simeq \sqrt{\frac{2\pi h}{A''(x_c)}}e^{-\frac{A(x_c)}{h}}
\end{equation}
to calculating the the higher order correction, in similar way yields 
\begin{equation}
I(h)=\sqrt{\frac{2\pi h}{A''(x_c)}}e^{-\frac{A(x_c)}{h}}\Upsilon(h)
\end{equation}
Which $\Upsilon$ given by\citep{2.0} 
\begin{eqnarray*}
\Upsilon(h) &&=1-\frac{h}{24}A^{(4)}\langle y^4 \rangle +\frac{h}{2\times6^2}{A'''}^2\langle y^6 \rangle + O(h^2)\\
&&=1+\frac{h}{24}\bigg(5 \frac{{A'''}^2}{{A''}^3}-3 \frac{A^{(4)}}{{A''}^2}\bigg)+O(h^2)
\end{eqnarray*}

\section{Double well and WKB}

There is an alternative way to solving the Schrödinger equation by WKB method, in this approach we consider wave function is given by the form of 
\begin{equation}
\psi(x)=e^{\frac{if(x)}{\hbar}}
\end{equation}
where $f(x)$ in general can be complex function and is possible to extend it by $\hbar$
\begin{equation}
f=f_0(x)+\hbar f_1(x) + {\hbar}^2 f_2(x)+\cdots
\end{equation}  

By putting the ansatz in the Schrödinger equation we have 
\begin{equation}
\frac{d^2}{dx^2}\psi(x)+\frac{p^2}{{\hbar}^2}\psi(x)
\end{equation}
and thus 
\begin{equation}
i\hbar f''-{f'}^2+p^2=0
\end{equation}
by using the explanation series for $f$ and collecting term in term of ${\hbar}^k , k=0,1,2,\cdots$ it yields 

\begin{equation}
{f_0'}^2=p , \ \ \ \ \ \ i{f_0''}=2f_0' f_1' , \ \ \ \ \ \ \ \ i f_1''=2f_0'f_2'+{f_1'}^2 , \cdots 
\end{equation}
if we keep the equations just to power of ${\hbar}^{+1}$ we have 
\begin{equation}
f_0=\pm \int p(x)dx+c'
\end{equation}
and
\begin{equation}
\frac{d f_1}{dx}=\frac{i}{2}\frac{{f_0}''}{{f_0}'}=\frac{i}{2}\frac{p'}{p} \rightarrow f_1=\frac{i}{2} \ln p +c''
\end{equation}
thus 
\begin{equation}
f=\pm \int p(x)dx + \hbar \frac{i}{2} \ln p+O({\hbar}^2)
\end{equation}
And the general form of wave function in WKB method can be obtain 
\begin{equation}
\psi \sim\frac{1}{\sqrt{p}}e^{\pm \frac{i}{\hbar}\int p dx }
\end{equation}

this estimation get intuition that WKB method provide answer by order $\hbar$ .

as usual we create the wave function by the WKB's recipe. The general form can be written $x>x_2$ by 

\begin{equation}
{\psi(x)}_{WKB}=\frac{D}{\sqrt{|p(x)|}}e^{-\frac{1}{\hbar}\int_{x_2}^{x}|p(x')|dx'} 
\end{equation}
and for the region $x_1<x<x_2$
\begin{equation}
{\psi}_{WKB}=\frac{1}{\sqrt{p(x)}}[Be^{\frac{i}{\hbar}\int_{x}^{x_2}p(x')dx'}+Ce^{-\frac{i}{\hbar}\int_{x}^{x_2}p(x')dx'}] 
\end{equation}

and similarly for $0<x<x_1$
\begin{equation}
{\psi}_{WKB}=\frac{1}{\sqrt{|p(x)|}}[Fe^{\frac{1}{\hbar}\int_{x}^{x_1}|p(x')|dx'}+Ge^{-\frac{1}{\hbar}\int_{x}^{x_1}|p(x')|dx'}] 
\end{equation}

Now for finding the relation between the  coefficient we need to introduce the patch-functions which describe our wave function near the turning points. If we describe the wave function around $x_2$by Airy function we have 
\begin{equation}
\psi(x)_{WKB}\approx \left\{
\begin{array}{rl}
\frac{D}{\sqrt{|p(x)|}}e^{-\frac{1}{\hbar}\int_{x_2}^{x} |p(x')|dx' } & x>x_2 \\
\frac{2D}{\sqrt{p(x)}}\sin [\frac{1}{\hbar}\int_{x}^{x_2}p(x')dx'+\frac{\pi}{4}] & x<x_2
\end{array} \right.
\end{equation} 

Therefor we write the wave function in the region $x_1<x<x_2$ by 
\begin{equation}
{\psi}_{WKB}\simeq\frac{2D}{\sqrt{p(x)}}\sin [\frac{1}{\hbar}\int_{x}^{x_2}p(x')dx'+\frac{\pi}{4}]
\end{equation}
if we rewrite it, we have 
\begin{equation}
{\psi}_{WKB}\simeq \frac{2D}{\sqrt{p(x)}}\sin[\frac{1}{\hbar} \int_{x_1}^{x_2} p(x')dx'-\frac{1}{\hbar}\int_{x_1}^{x}p(x')dx'+\frac{\pi}{4}]
\end{equation}
Introducing this parameter is useful
\begin{equation}
\theta:=\int_{x_2}^{x_1} p(x')dx'
\end{equation}
using from this definition yields
\begin{equation}
{\psi}_{WKB}\simeq -\frac{2D}{\sqrt{p(x)}}\sin[\frac{1}{\hbar}\int_{x_1}^{x}p(x')dx'-\theta -\frac{\pi}{4}] , x_1<x<x_2
\end{equation}
Again if we analyse the relations around $x_1$ point , for the region $x<x_1 $

\begin{equation}
\psi(x)_{WKB}\simeq \frac{1}{\sqrt{|p(x)|}}[F e^{\frac{1}{\hbar}\int_{x}^{x_1}|p(x')|dx'}+G e^{-\frac{1}{\hbar} \int_{x}^{x_1}|p(x')|dx'}]
\end{equation}
and for the $x>x_1$
\begin{equation}
\psi(x)_{WKB}\simeq  -\frac{2D}{\sqrt{p(x)}}\sin[\frac{1}{\hbar}\int_{x_1}^{x}p(x')dx'-\theta -\frac{\pi}{4}]
\end{equation}

now we calculate the $p(x)$ and write the Airy function for the lef hand side of the $x_1$ which has negative slope 
\begin{equation}
{\psi}_{WKB}\simeq \frac{1}{\sqrt{\hbar} {\alpha}^{\frac{3}{4}}(-x)^{\frac{1}{4}}}[F e^{\frac{2}{3}(-\alpha x)^{\frac{3}{2}}}+Ge^{-\frac{2}{3}(-\alpha x)^{\frac{3}{2}}}]
\end{equation}
\begin{equation}
{\psi}_{p} \simeq \frac{a}{2\sqrt{\pi}(-\alpha x)^{\frac{1}{4}}}e^{-\frac{2}{3}(-\alpha x)^{\frac{3}{2}}}+\frac{a}{\sqrt{\pi}(-\alpha x)^{\frac{1}{4}}}e^{\frac{2}{3}(-\alpha x)^{\frac{3}{2}}}
\end{equation}
here we considered $x_1=0$ for  simplicity
, we turning it to it’s original value later,by  comparing the answers we have
\begin{equation}
a=2G\sqrt{\frac{\pi}{\hbar \alpha}} , \ \ \ \ \ \ \ \ b=F\sqrt{\frac{\pi}{\hbar \alpha}}
\end{equation}
now we do this algorithm for the the other side, around $x_1$, but here the slpoe is positive
\begin{equation}
{\psi}_{WKB}\simeq -\frac{2D}{\sqrt{\hbar}{\alpha}^{\frac{3}{4}}x^{\frac{1}{4}}} \sin[\frac{2}{3}(\alpha x)^{\frac{3}{2}}-\theta -\frac{\pi}{4}]
\end{equation}
\begin{equation}
{\psi}_p \simeq \frac{a}{\sqrt{\pi}(\alpha x)^{\frac{1}{4}}}\sin[\frac{2}{3}(\alpha x)^{\frac{3}{2}}+\frac{\pi}{4}]+\frac{b}{\sqrt{\pi}(\alpha x)^{\frac{1}{4}}}\cos[\frac{2}{3}(\alpha x)^{\frac{3}{2}}+\frac{\pi}{4}]
\end{equation}

For better understanding we extend the sine and cosine in the previous relations, which yields 
\begin{equation}
{\psi}_{WKB} \simeq -\frac{2D}{\sqrt{\hbar}{\alpha}^{\frac{3}{4}}}\frac{1}{2i}[e^{i\frac{2}{3}(\alpha x)^{\frac{3}{2}} e^{-i\theta}e^{-i\frac{\pi}{4}}}+e^{-i\frac{2}{3}(\alpha x)^{\frac{3}{2}} e^{i\theta}e^{i\frac{\pi}{4}}}]
\end{equation}
\begin{eqnarray}
{\psi}_p \simeq && \frac{1}{\sqrt{\pi}{\alpha}^{\frac{1}{4}}} \frac{a}{2i}[ e^{i\frac{2}{3}(\alpha x)^{\frac{3}{2}}}e^{i\frac{\pi}{4}}-e^{-i\frac{2}{3}(\alpha x)^{\frac{3}{2}}}e^{-i\frac{\pi}{4}} ] \nonumber \\
&&+\frac{b}{2}[e^{i\frac{2}{3}(\alpha x)^{\frac{3}{2}}}e^{i\frac{\pi}{4}}+e^{-i\frac{2}{3}(\alpha x)^{\frac{3}{2}}}e^{-i\frac{\pi}{4}}]
\end{eqnarray}
therefor 
\begin{eqnarray}
&&-2D \sqrt{\frac{\pi}{\alpha \hbar}}e^{-i\theta}e^{-i\frac{\pi}{4}}=(a+ib)e^{i\frac{\pi}{4}}\\
&& 2D \sqrt{\frac{\pi}{\alpha \hbar}}e^{i\theta}e^{i\frac{\pi}{4}}=(-a+ib)e^{-i\frac{\pi}{4}}
\end{eqnarray}
or 
\begin{equation}
(a+ib)=2D \sqrt{\frac{\pi}{\alpha \hbar}}i e^{-i\theta} , (a-ib)=-2D \sqrt{\frac{\pi}{\alpha \hbar}}i e^{i\theta}
\end{equation}
thus 
\begin{equation}
a=2D \sqrt{\frac{\pi}{\alpha \hbar}}\sin{\theta} 
\end{equation}
\begin{equation}
b=2D \sqrt{\frac{\pi}{\alpha \hbar}}\cos{\theta} 
\end{equation}
now we can express $F$ and $G$ by $D$ 
\begin{equation}
G=D\sin{\theta} ,\ \ \ \ \ \ \ \ \ F=2D\cos{\theta}
\end{equation}
%

now, as conclusion, for $0<x<x_1$
\begin{equation}
\psi(x)\simeq \frac{D}{\sqrt{|p(x)|}}[2\cos{\theta}e^{\frac{1}{\hbar}\int_{x}^{x_1}|p(x')dx'}+\sin{\theta} 
 e^{-\frac{1}{\hbar}\int_{x}^{x_1}|p(x')|dx'}] 
\end{equation}

and for $x_1<x<x_2$

\begin{equation}
\psi(x)\simeq  \frac{2D}{\sqrt{p(x)}}\sin[\frac{1}{\hbar}\int_{x}^{x_2}p(x')dx'+\frac{\pi}{4}]  
\end{equation}
and $x>x_2$
\begin{equation}
\psi(x)\simeq  \frac{D}{\sqrt{|p(x)|}}e^{-\frac{1}{\hbar}\int_{x_2}^{x}|p(x')|dx'} 
\end{equation}
%
Now we focus on finding the spectrum of energy in this system as the potential in symmetric 
$V(x)=V(-x) $ 
The Eigen functions of Hamiltonian can have completely positive or negative parity, if the wave function be odd then it must 
$\psi(x=0)=0$ 
Therefor
\begin{equation}
2\cos{\theta}e^{\frac{1}{\hbar}\int_{x}^{x_1}|p(x')dx'}+\sin{\theta} e^{-\frac{1}{\hbar}\int_{x}^{x_1}|p(x')|dx'}=0
\end{equation}
if we define 
\begin{equation}
\frac{1}{\hbar} \int_{0}^{x_1}|p(x')dx'=\frac{1}{2}\phi 
\end{equation}
Then
\begin{equation}
\tan{\theta}=-2e^{\phi}
\end{equation}
in the other hand if the wavefunction be even then it must
${\psi}'(x=0)=0$ 
Therefor 
\begin{eqnarray}
&&-\frac{1}{2} \frac{D}{(|p(x)|)^{\frac{3}{2}}}\frac{d|p(x)|}{dx}|_0[2\cos{\theta}e^{\frac{\phi}{2}}+\sin{\theta} e^{-\frac{\phi}{2}}]
\nonumber
\\
&&+
\frac{D}{\sqrt{|p(x)|}}[2\cos{\theta}e^{\frac{1}{\hbar} \int_{0}^{x_1} |p(x')|dx'}(-\frac{1}{\hbar}|p(0)|)
\nonumber 
\\
&&+\sin{\theta}e^{\frac{1}{\hbar} \int_{0}^{x_1} |p(x')|dx'}(-\frac{1}{\hbar}|p(0)|)]=0
\end{eqnarray}
hence 
$\frac{d|p(x)|}{dx}=\frac{d}{dx}\sqrt{2m[V(x)-E]}=\sqrt{2m}\frac{1}{2}\frac{1}{\sqrt{V-E}}\frac{dV}{dx}$ and $ \frac{dV}{dx}|_0=0$
therefor we find that 
\begin{equation}
2\cos{\theta}e^{\frac{\phi}{2}}=\sin{\theta}e^{-\frac{\phi}{2}}
\end{equation}
or in the equivalent way 
\begin{equation}
\tan{\theta}=+ 2e^{\phi}
\end{equation}
Thus, in general we have 
\begin{equation}
\tan{\theta}=\pm 2e^{\phi} , 
\end{equation}
Which $\theta$ and $\phi$ were defined before .
This relation describe the approximated spectrum of enregy.it is important to notice that the $x_1$ and $x_2$ are themself function of $E$ therefore $\theta$ and $\phi$ are function of $E$.
it is possible to discuses more on result if we consider the well penetration be long or long , in this approximation regime 
$e^{\phi}$
has high value and therefore 
$\theta$ in near to half-number of $\pi$.thus 
\begin{eqnarray}
\tan{\theta}&&=\tan[(n+\frac{1}{2})\pi+\epsilon]=\frac{\sin[(n+\frac{1}{2})\pi+\epsilon]}{\cos[(n+\frac{1}{2})\pi+\epsilon]}\nonumber \\
&&=\frac{(-1)^n\cos{\epsilon}}{(-1)^{n+1}\sin{\epsilon}}=-\frac{\cos{\epsilon}}{\sin{\epsilon}}\approx -\frac{1}{\epsilon}
\end{eqnarray}
this yields
\begin{equation}
\epsilon \simeq \mp \frac{1}{2}e^{-\phi} \rightarrow \theta-(n+\frac{1}{2})\pi \approx \mp \frac{1}{2}e^{-\phi}
\end{equation}
thus 
\begin{equation}
\theta \simeq (n+\frac{1}{2})\pi \mp \frac{1}{2}e^{-\phi}
\end{equation}

In practice it means that 
\begin{eqnarray}
&&\frac{1}{\hbar}\int_{x_1}^{x_2} \sqrt{2m[E-V(x')]}dx' \nonumber \\
&&\simeq(n+\frac{1}{2})\pi \mp \frac{1}{2}e^{-\frac{1}{\hbar}\int_{-x_1}^{x_1}\sqrt{2m[V(x')-E]}dx'}
\end{eqnarray}

For instance if our potential has the symmetric shape like $x^2$ or begin more specific be like 
$V(x)=\frac{1}{2} m {\omega}^2(x+a)^2 , x<0 $ and$ \frac{1}{2}m{\omega}^2(x-a)^2 , x>0$ 
Then simple calculation can shows that 
\begin{equation}
E_{n}^{\pm} \simeq (n+\frac{1}{2})\hbar \omega \mp \frac{\hbar \omega}{2\pi}e^{-\phi}
\end{equation}
Which by defining $z:=\sqrt{\frac{m}{2E}}\omega (a-x)$ we have 

\begin{equation}
\phi=\frac{2E}{\hbar \omega}[z_0\sqrt{z_0^2-1}-\ln(z_0+\sqrt{z_0^2-1})]
\end{equation}



\end{appendices}


\begin{thebibliography}{99}
\bibitem{101} Manin, Yu. I. \textit{Vychislimoe i nevychislimoe [Computable and Noncomputable] (in Russian)}, Sov.Radio. pp. 13–15,1980 

\bibitem{102} Feynman, R. P.\textit{ Simulating physics with computers}, International Journal of Theoretical Physics\textbf{ 21}  467–488,1982 .


\bibitem{103}  Simon, D.R.  ,\textit{ On the power of quantum computation},  Foundations of Computer Science,  1994
\bibitem{104}  Nielsen, Michael A., Chuang, Isaac L. ,\textit{ Quantum Computation and Quantum Information},Cambridge University Press,2001

\bibitem{105} Bennett C.H., Bernstein E., Brassard G., Vazirani U, \textit{The strengths and weaknesses of quantum computation}. SIAM Journal on Computing \textbf{26}: 1510–1523 ,1997.


\bibitem{106}  M. A. Nielsen and I. L. Chuang,\textit{ Quantum Computation
and Quantum Information} ,Cambridge, 2000.


\bibitem{111}   J. Q.You ,Franco Norim\textit{Superconducting Circuits and Quantum Information}, Physics Today,November, 2005
\bibitem{107}Y. Makhlin, G. Sch¨on and A. Shnirman,\textit{ Rev. Mod. Phys.},
\textbf{73}, 357 ,2001.

\bibitem{108}G. Wendin and V. S. Shumeiko, \textit{Fiz. Nizk. Temp.} 
 [Low Temp. Phys. ],33, 724 ,2007.
\bibitem{109} J.Q. You and F. Nori,\textit{ Phys. Today },text\textbf{58}, 42,2005.
\bibitem{110} J. Clarke and F. K. Wilhelm, \textit{Nature},\textbf{ 453}, 1031, 2008.
\bibitem{112}  A.Blais  \textit{Superconducting qubit systems come of age}, Physics today, July, 2009
,
\bibitem{201}  P. Dirac,\textit{The Lagrangian in Quantum Mechanics}, Physikalische Zeitschrift der Sowjetunion \textbf{3} 64–72,1933

\bibitem{202} Masud Chaichian, Andrei Pavlovich Demichev , \textit{ Path Integrals in Physics}  Volume 1: Stochastic Process and  Quantum Mechanics. Taylor and  Francis. p. 1 ff. ISBN 0-7503-0801-X,2001

\bibitem{1}     C.Coleman, \textit{Aspect of symmetry: selected Erice Lectures of Sidney Coleman},
    Cambridge University Press, Cambridge , 1985.
    
\bibitem{2}     Alexander Atland and Ben Simons , second edition \textit{cindensed Matter Field Theory} ,Cambridge University Press,2010 .
\bibitem{2.0}   J.Zinn-Justin, \textit{Path Integrals in Quantum Mechanis },Oxford University Press ,1993.
\bibitem{2.2}   R.Feynman and A.Hibbs, \textit{Quantum Mechanics and Path Integrals} ,McGraw-Hill,New Yourk , 1965.
\bibitem{12}    A. M.Polyakov, \textit{Nucl. Phys} \textbf{B121}, 429 (1997).
\bibitem{13}    J. L. Gervais and B.Sakita, \textit{Phys .Rev .} \textbf{D11}, 2942 (1975).
\bibitem{14}    C. Callan and S.Coleman , \textit{Phys .Rev . } \textbf{D16},1762(1997).
    
\bibitem{15}    J. S. Lsnger, \textit{Ann. Phys.(N.Y)} \textbf{41}, 108 (1967).
\bibitem{16}    L.R. Schulman,  
 \textit{Techniques and Applications of Path Integration} ,Wiley,New York,1981.

\bibitem{18}    A. M. Polyakov,  \textit{Gauge Fields and Strings}, Harwood, 1987.
\bibitem{19}    A. M. Polyakov, \textit{Quark confinement and topology}, \textit{Nucl. Phys.}, \textbf{B120} 429-58(1977).

\bibitem{20}    P. W. Anderson, B. I. Halperin, and C. M. varma, \textit{Anomalous low-temperature thermal properties of glasses and spin glasses,Phil. Mag. } \textbf{25},1-9, 1972

\bibitem{21}    H. Kleinert,  \textit{Path Integrals in Quantum Mechanics, Statistics and  Polymer Physics},World Scientific, Singapore, 1995


\bibitem{22}   C. Grosche and F. Steiner, \textit{Handbook of Feynman path integrals }, Springer,Berlin, Heidelberg, 1998.
\bibitem{23}   L.D. Faddeev, \textit{in Methods in Field Theory}, Les Houches School ,1975.
\bibitem{24}   R.G. Newton, \textit{Scattering theory
Waves and Particles },McGraw-Hill, New York ,1966.
\bibitem{25} G. Wendin and V.S. Shumeiko,\textit{ superconducting  quantum circuits, Qubits and computing wendin review },2005 
\bibitem{30} Hesam Zandi, Shabnam Safaei, Sina Khorasani, Mehdi Fardmanesh, \textit{Study of Junction and Bias Parameters in Readout of Phase Qubits},  Physica C: Superconductivity and its applications , DOI: 10.1016/j.physc.2011.05.002, 2012.
\end{thebibliography}
\end{document}